\documentclass[a4paper, 10pt, journal]{IEEEtran}

\usepackage{url}
\usepackage{graphicx}
\usepackage{amsfonts}
\usepackage[shortlabels]{enumitem}
\usepackage{color}
\usepackage{stfloats}
\usepackage{color,soul}
\usepackage{amsmath}
\usepackage{multirow}
\usepackage{hhline}
\usepackage{url}
\usepackage{float}
\usepackage{enumitem}
\usepackage{cite}
\usepackage{multicol}
\usepackage{color, colortbl}
\usepackage{blindtext}
\usepackage{rotating}
\usepackage{array}
\usepackage{tikz}
\definecolor{Gray}{gray}{0.9}
\usepackage{booktabs}
\usepackage{graphicx}
\usepackage[normalem]{ulem}
\usepackage{xltabular}
\usepackage{lscape}
\usepackage{hyperref}
\usepackage{makecell}

\usepackage{url}

\usepackage{breakurl}

\usepackage[acronym,nogroupskip,nonumberlist]{glossaries}
\useunder{\uline}{\ul}{}
\makeglossaries

\newacronym{mpc}{MPC}{model predictive control}
\newacronym{AI}{AI}{Artificial Intelligence}
\newacronym{ML}{ML}{Machine Learning}
\newacronym{DL}{DL}{Deep Learning}
\newacronym{BS}{BS}{Base Station}
\newacronym{RL}{RL}{Reinforcement Learning}
\newacronym{DRL}{DRL}{Deep Reinforcement Learning}
\newacronym{TLA}{TLA}{Three Letter Acronym}
\newacronym{5G}{5G}{Fifth Generation}
\newacronym{6G}{6G}{Sixth Generation}
\newacronym{B5G}{B5G}{Beyond Fifth Generation}
\newacronym{SLA}{SLA}{Service Level Agreement}
\newacronym{E2E}{E2E}{End to End}
\newacronym{SDN}{SDN}{Software Defined Networks}
\newacronym{NFV}{NFV}{Network Function Virtualization}
\newacronym{VNF}{VNF}{Virtual Network Functions}
\newacronym{RAN}{RAN}{Radio Access Network }
\newacronym{IoT}{IoT}{Internet of the Things}
\newacronym{XAI}{XAI}{Explainable Artificial Intelligence}
\newacronym{LIME}{LIME}{Local Interpretable Model-Agnostic Explanations}
\newacronym{SHAP}{SHAP}{Shapley Additive Explanations}
\newacronym{CFE}{CFE}{Counterfactual Explanations}
\newacronym{LRP}{LRP}{Layer-wise Relevance Propagation}
\newacronym{PIRL}{PIRL}{Programmatically Interpretable Reinforcement Learning framework }
\newacronym{LMUT}{LMUT}{ Linear Model U-Trees}
\newacronym{IBN}{IBN}{Intent Based Network}
\newacronym{DLT}{DLT}{Distributed Ledger Technology}
\newacronym{VR}{VR}{Virtual Reality}
\newacronym{AR}{AR}{Augmented Reality}
\newacronym{XR}{XR}{Extended Reality}
\newacronym{MR}{MR}{Mixed Reality}
\newacronym{MD}{MD}{Management Domain}
\newacronym{UE}{UE}{User Equipment}
\newacronym{DOS}{DoS}{Denial of Services}
\newacronym{IDS}{IDS}{Intrusion Detection Systems}
\newacronym{UAV}{UAV}{Unmanned Aerial Vehicles}
\newacronym{CPS}{CPS}{Cyber Physical Systems}
\newacronym{RIC}{RIC}{RAN Intelligence Controller}
\newacronym{MLP}{MLP}{Multi Layer Perceptrons}
\newacronym{SVM}{SVM}{Support Vector Machines}
\newacronym{CRAN}{CRAN}{Centralized/Cloud Radio Access Network}
\newacronym{VRAN}{VRAN}{Virtualized Radio Access Network}
\newacronym{ORAN}{ORAN}{Open Radio Access Network}
\newacronym{DDOS}{DDoS}{Distributed Denial of Service}
\newacronym{GDPR}{GDPR}{General Data Protection Regulation}
\newacronym{MITM}{MITM}{Man in the Middle}
\newacronym{PUEA}{PUEA}{Primary User Emulation Attack}
\newacronym{FL}{FL}{Federated Learning}
\newacronym{IIOT}{IIoT}{Industrial Internet of Things}
\newacronym{DQN}{DQN}{Deep Q-Network}
\newacronym{DNN}{DNN}{Deep Neural Network}
\newacronym{ZSM}{ZSM}{Zero-touch network and Service Management}
\newacronym{VAE}{VAE}{Variational Auto-encoders}
\newacronym{PCA}{PCA}{Principal Component Analysis}
\newacronym{SFIT}{SFIT}{Single Feature Introduction Test}
\newacronym{DP}{DP}{Differential Privacy}
\newacronym{HE}{HE}{Homomorphic Encryption}
\newacronym{MPC}{MPC}{Multi-Party computation}
\newacronym{TEE}{TEE}{Trusted Execution Environment}
\newacronym{KPM}{KPM}{Key Performance Measurements}
\newacronym{DPI}{DPI}{Deep Packet Inspection}
\newacronym{CPE}{CPE}{Customer Premises Equipment}
\newacronym{RIS}{RIS}{Reconfigurable Intelligent Surfaces}
\newacronym{LLMs}{LLMs}{Large Language Models}
\newacronym{CoT}{CoT}{Chain-of-Thought}

\newglossarystyle{mystyle}
{
  {
    \begin{center}
    \begin{tabular}{p{1.5cm}p{\glsdescwidth}}
    \hline
    \rowcolor{gray!30}
    \textbf{Acronym} & \textbf{Definition}\\
    \hline
    \midrule

    }
  {
     \bottomrule
     \end{tabular}
     \end{center}
  }%
  \renewcommand{\glossarysection}[2][]{}
  \renewcommand*{\glsgroupheading}[1]{}%
}

\glsdisablehyper

\begin{document}

\title{A Survey on XAI for 5G and Beyond Security: Technical Aspects, Challenges and Research Directions}

\author
{Thulitha Senevirathna, 
Vinh Hoa La, Samuel Marchal, Bartlomiej Siniarski,       
 ~Madhusanka~Liyanage,~\IEEEmembership{Senior Member,~IEEE},
 and Shen Wang, ~\IEEEmembership{Senior Member,~IEEE}

\thanks{Thulitha Senevirathna is with the School of Computer Science, University College Dublin, Ireland, email: thulitha.senevirathna@ucdconnect.ie }
\thanks{Vinh Hoa La is with Montimage, France, email: vinh\_hoa.la@montimage.com }
\thanks{Samuel Marchal is with VTT Technical Research Centre of Finland, email: samuel.marchal@vtt.fi}
\thanks{Bartlomiej Siniarski is with the School of Computer Science, University College Dublin, Ireland, email: bartlomiej.siniarski@ucd.ie}
\thanks{Madhusanka Liyanage is with the School of Computer Science, University College Dublin, Ireland email: madhusanka@ucd.ie}
\thanks{Shen Wang is with the School of Computer Science, University College Dublin, Ireland, email: shen.wang@ucd.ie}
}

\maketitle

\begin{abstract}

With the advent of 5G commercialization, the need for more reliable, faster, and intelligent telecommunication systems is envisaged for the next generation beyond 5G (B5G) radio access technologies. Artificial Intelligence (AI) and Machine Learning (ML) are immensely popular in service layer applications and have been proposed as essential enablers in many aspects of 5G and beyond networks, from IoT devices and edge computing to cloud-based infrastructures. However, existing 5G ML-based security surveys tend to emphasize AI/ML model performance and accuracy more than the models' accountability and trustworthiness. In contrast, this paper explores the potential of Explainable AI (XAI) methods, which would allow stakeholders in 5G and beyond to inspect intelligent black-box systems used to secure next-generation networks. The goal of using XAI in the security domain of 5G and beyond is to allow the decision-making processes of ML-based security systems to be transparent and comprehensible to 5G and beyond stakeholders, making the systems accountable for automated actions. In every facet of the forthcoming B5G era, including B5G technologies such as ORAN, zero-touch network management, and end-to-end slicing, this survey emphasizes the role of XAI in them that the general users would ultimately enjoy. Furthermore, we presented the lessons from recent efforts and future research directions on top of the currently conducted projects involving XAI.

\end{abstract}


\section{Introduction} \label{sec:intro} 

The wireless communication industry is one of the most rapidly developing sectors in technology. The innovations that thrive in the telecommunication sector have laid the infrastructure and led towards a consonant development that has led to exponential growth in living standards. The first generation of cellular networks started evolving wireless communication technology in the 1980s. 5G wireless technology, primarily based on softwarization, is expected to complete the transition with significant coverage by 2025. The most noticeable feature of 5G is the cloudification of networks via microservices-based architecture. With the start of commercialized implementation of 5G, experts predict that 6G mobile communication will become widely available in the following years \cite{porambage2021roadmap}.
Meanwhile, the academic community is more focused on new lines of study in advance of the beyond 5G or 6G standardization. Edge intelligence (EI), beyond 6GHz to THz communication, non-Orthogonal Multiple Access (NOMA), \gls{RIS}, and Zero-touch Networks have risen in recent years \cite{de2021survey, benzaid2020zsm, saad2019vision}. While allowing exceptionally high data rates potentially reaching tens or hundreds of gigabits per second, THz communication has a shorter range than mmWaves in 5G communication. In B5G, enablers such as RIS alleviate blockage vulnerability and enhance coverage for THz communication. These concepts are being developed into the technology that will power the next generation of communication networks. There is still a long way to go in terms of 5G network capabilities to meet the needs of these applications, which need high-speed data transfer rates and real-time access to vital computing resources. IoE, enabled by 5G, seeks to connect vast numbers of devices and \gls{CPS}, surpassing 5G's capabilities into the B5G era. For example, 6G is expected to connect millions of devices and provide instant access to tremendous computing and storage capabilities. For B5G wireless networks, the scientific community expects fully intelligent network orchestration and management \cite{de2021survey, you2021towards}. It will be distinct from previous generations in various aspects, including network infrastructures, radio access methods, processing and storage capacities, and application types. New applications must intelligently use communications, compute, control, and storage resources. Moreover, wireless networks are producing a large amount of data. This paradigm shift allows data-driven real-time network design and operation in 5G and beyond.

Physical attacks, eavesdropping, and authentication and authorization issues plagued wireless communication technologies from 1G to 3G. It now includes more complicated attacks and tougher assailants. Many security improvements came to fruition with 4G. However, with the larger landscape of connectivity points, an increase in the potential for security loopholes is inevitable. For example, the 4G core network is vulnerable to DoS attacks \cite{paolini2012wireless}. Spam over Internet Telephony (SPIT), which is spam for VoIP, spoofing, where an attacker misdirects the users with fraudulent data, and SIP registration hijacking, where IP packet headers are replaced with attacker's ones, are some of the possible threats 4G \cite{park2007survey}. These attacks have morphed into \gls{SDN}, \gls{NFV} and cloud computing in the 5G. Insecure \gls{SDN} features include OpenFlow, centralized network administration (prone to DoS attacks), core and backhaul, edge device vulnerabilities, and open APIs \cite{liyanage2016opportunities, liyanage2018comprehensive}. Research communities are starting to focus on security vulnerabilities in 5G communication using advanced networking, \gls{AI}/\gls{ML}, and linked intelligence technologies that power the B5G vision. On top of the unsolved security issues brought forward by previous generations, these new technologies open 5G and beyond networks to a whole new threat surface that has never been seen before. Nevertheless, the overall success of B5G ultimately depends on how well AI and 6G cooperate in the future \cite{schneier2018artificial}.

AI changes the threat landscape and constraints on potential applications before they see the light of day. The at-risk complex systems include smart \gls{CPS} (SCPSs). It is important to note that the interconnectivity of SCPSs is rapidly increasing with the aid of \gls{IoT}, \gls{AI}, Wireless Sensor Networks (WSNs), and cloud computing. This interconnectivity is the backbone for a vast array of services and applications in the 5G and beyond\cite{kaloudi2020ai}. A single vulnerability in SCPSs can cause catastrophic failures (Butterfly effect) due to their intertwined nature. This characteristic of SCPSs could give rise to larger-scale attacks, unlike those observed before the advent of 5G.

As a consequence, all interconnected devices and users stand at risk. Even though research on AI to protect against cyber threats has been ongoing for many years \cite{li2017machine, marino2018adversarial}, it is still unclear how to ensure the security of networks with AI integrated into their core operations. A significant drawback in AI security has derived from the black-box nature of those systems in one way or the other. Therefore, maintaining accountable and trustworthy AI in this regard is highly important.

The Defense Advanced Research Projects Agency (DARPA) started the \gls{XAI} initiative in May 2017 to develop a set of new AI methodologies that would allow end-users to comprehend, adequately trust, and successfully manage the next generation of AI systems \cite{gunning2019darpa}. To further elaborate, it is a collective initialization of computer sciences and the social sciences, which includes human psychology of explanations. The overall success of 5G and beyond would ultimately stand on how far the \gls{AI} used in its implementation is going to be resilient and trustworthy for the general public for utilization \cite{schneier2018artificial}. Extending research on possible techniques such as XAI in this regard is a crucial step that needs to be taken abruptly.  

\begin{figure*}[htb]
    \centering
    \includegraphics[width=1\textwidth]{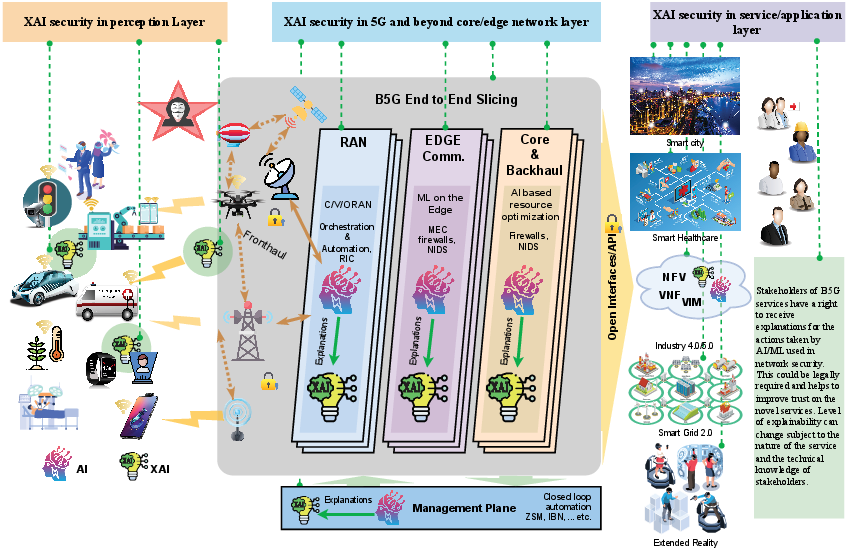} 
    \vspace{1mm}
    \caption{XAI is applicable in many facets of each layer in the 3-layered 5G and beyond architecture. XAI methods deployed around security algorithms in the perception layer would enhance the interpretability of the devices and contain additional information as they are the closest accessible points for the general users. The ubiquitous use of ML applications in the network layer requires quantifiable approaches to interpretability. In the network layer, XAI will become an essential component in the interactions between operators and the ML model. Interpretations generated in the first and second layers approach the users through the service layer. The comprehensiveness and relevance of the explanations will determine the attraction of new clients and the retention of existing clients for service providers.
    }
    \label{fig:xai_b5g}
\end{figure*}

\begin{table}[h!]
\centering
\caption{Summary of Important Acronyms}
\label{table:acronyms}
\printglossary[type=\acronymtype,style=mystyle]
\end{table}

\subsection{Paper Motivation}

When writing this article, 5G is commercially rolling out, with many researchers focusing on the B5G. Its applications, architecture, and enabling technologies are the subject of many recent studies published, as shown in Table \ref{tab:surveys}. In addition, studies such as \cite{shafin2020artificial, zhao2020comprehensive,  chowdhury20206g, saad2019vision, dang2020should, david20186g, chen2020vision, letaief2019roadmap, de2021survey, jiang2021road } have mainly focused on the vision, potential applications and requirements of the B5G wireless communication technologies such as terabits per second speeds FeMBB, connected intelligence, and EDuRLLC, among others that would facilitate up and coming applications such as autonomous vehicles, telemedicine, the extended reality in the future.

Among key enablers in B5G/6G mobile communication, such as THz communication, edge computing, swarm networks, full automation, and blockchain, \gls{AI} takes a prominent place. AI techniques are more suitable for solving complex problems due to their generalization capabilities and, thus, are fit for use in many novel B5G-era applications. Studies including \cite{ sheth2020taxonomy, shafin2020artificial, kato2020ten, xiao2020towards, siriwardhana2021ai, porambage2019sec, chen2020deep}, elaborate on the importance of AI and its trends in B5G, and the challenges it brings to future communication technologies. Previous surveys such as \cite{ liyanage2016opportunities, nguyen2021security, gui20206g, millar2019intelligent, kumar2014survey, tian2017survey } highlight the dynamics of security aspects in a range of 5G and beyond enabling technologies such as IoT, RAN and edge computing, while \cite{ schneier2018artificial, millar2019intelligent, caroline2020artificial, barreno2006can } focus entirely on the security threats and potential defences that would improve the trust in AI/ML methods used in 5G and beyond. 

Although it shows promising results, only a few publications (\cite{guo2020explainable, arrieta2020explainable, das2020opportunities }) have covered the XAI applications in the context of security or XAI research projects and standardization methods. Opportunities, challenges, and standardizations in XAI are still in their infancy, and more collaborative work is needed with experts from fields such as human psychology and sociology to move towards more concrete real-world applications. Summarized table \ref{tab:surveys} outlines contemporary research and surveys conducted on the advancements of B5G, AI, and XAI. Here, we have found that each paper presents applications in disarticulated contexts. On the contrary, implementing 5G and beyond technologies begs for a holistic review of AI and XAI in security, given that accountability and resilience are core and essential characteristics of any mobile network generation. Many researchers focus on B5G, XAI, and AI techniques in sunders, but currently, there has not been a cognate approach where the viability of XAI techniques has been reviewed in the context of 5G and beyond. As a response, this survey reports a comprehensive overview of XAI and security technical aspects, applications, requirements, limitations, challenges/issues, current projects, standardization initiatives, and lessons learned for the beyond 5G applications.

\subsection{XAI for 5G and beyond: a data life cycle approach} \label{SECXAIFORB5G}

Explainable Artificial Intelligence (XAI) represents an advancement over the opaque AI systems in networking. Starting with the 5G era, artificial intelligence (AI) is anticipated to assume various roles across all levels of mobile networks. Furthermore, explainable AI (XAI) would be the subsequent phase in attaining accountability and transparency in AI systems. The architecture of 5G and future networks has to be reconfigured to fully accept this new paradigm of wireless AI architecture and its data life cycle. 

We propose to slightly modify the three-layer architecture for 5G and beyond, as shown in Fig.~\ref{fig:xai_b5g} as the basis of our approach towards XAI in future networks. The three-layered architecture is explicitly designed for IoT systems \cite{dai2019blockchain, guo2021enabling, nguyen2021security}. The three-layered architecture is based on the data flow of sensors and devices in the IoT era that aligns perfectly with the wireless AI architecture \cite{ nguyen2020wireless, zhang2023research, filippou2022pervasive} that rests at the core of XAI-driven 5G and beyond security. The data life cycle refers to how a system generates, collects, processes, and analyses data. In the context of the 5G and beyond networks, the three-layered architecture can be mapped almost perfectly to those operational aspects of AI-driven architecture. 
It's important to note that the OSI (Open Systems Interconnection) model and the three-layer architecture in IoT (Internet of Things) play a significant role in conceptualizing the structure and functionality of networked systems. While they serve different purposes and do not have direct one-to-one correspondence, we can draw parallels to understand their relationship. Our discussion of the three-layered architecture as the reference model for wireless data-driven XAI in the 5G and beyond network model is based on this understanding.  

\begin{itemize}
\item \textit{The perception layer: }This includes sensors, actuators, controllers, bar code/QR code tags, RFID tags, smart meters, and other wireless/wired devices. These versatile gadgets can detect and gather data from the environment, while some devices can act on the environment based on the data they receive. In this layer, the physical observations and measurements are transformed into digital data, spawning XAI's wireless data life cycle. It would also include pre-model explanations for security that are generated in the devices. The perception layer has no direct correspondence to one single OSI layer. However, it performs multiple operations that span across physical and data link layers (e.g., protocols: NFC, ZigBee, RFID) of the OSI model.

\item \textit{The network layer: } Enables data transmission, routing, and communication protocols, letting devices and sensors deliver data to the cloud or other processing points. The data would go through ORAN, backhaul, and core networks to reach their destinations. Also, with 5G and beyond fully virtualized future networks, the cloud and edge paradigms play a significant role in explanation generation and consumption. The second stage of the wireless data life cycle for AI happens mainly in the network layer. Although there is not one analogous layer for the network layer in IoT in the OSI model, it serves a heuristically similar purpose as the transport and network layers from the OSI model.

\item \textit{The service/application layer: } This layer is where industrial applications take place. For example, smart grid, industry 4.0/5.0 food industry, and smart health. In the wireless data life cycle concept, stakeholders, such as theorists (researchers), ethicists, XAI model creators, and end users, process and harvest useful information through X/AI models. The security of this layer is of utmost importance as it could decide users' confidence towards new technologies such as X/AI, depending on their accountability, transparency, and fairness. The application layer also involves data processing, analysis, and decision-making in 5G and beyond applications, heuristically similar to OSI application, session, and presentation layers.
\end{itemize}

Although the three-layered architecture could map the life cycle of wireless data-driven X/AI, it is also important to realize that there are exceptions to this. For example, the data can be collected by service providers for the maintenance of the networks directly from the physical measurements. It could include user/client surveys, reports, and cross-organizational data sharing \cite{nguyen2020wireless}. In the rest of the article, we will discuss the adaptability of XAI in the network layer to maintain cohesiveness to the 5G and beyond era of networks. It would enhance the reading experience and minimize the confusion that could arise due to the ubiquitous nature of the application possibilities of XAI. By narrowing down our scope to the security of concepts native to 5G and beyond future networks, we intend to provide the reader with a rich understanding of XAI's future research potential.

During the discussions pertaining to the above architecture, we use a 6W technique building upon the work of Vigano et al. (2020) \cite{vigano2020explainable}. Here, we advocate for assessing the 6W questions - Why, Who, What, Where, When, and How - as a means to produce comprehensive security explanations when creating a system with explainable security. Figure \ref{fig:xsec} depicts the flow of identifying basic building blocks to design an explainable security system. First, the apparent reason to \textit{why} the system needs XAI must be identified. Then \textit{to whom} and \textit{who} create the explanation and decide the granularity level of the content broadcast to each group of actors. Identifying the needs of each actor early on helps to decide on \textit{what} aspects of the system need to be explained. Here, the system designers must consider the layer of B5G architecture and fit the explanation to meet its requirements. Although the explanation is generated in one layer, it will not be the same \textit{where} it will be accessible. It must be decided whether it will be a separate service or embedded in the system/output. It is also essential to decide the \textit{when} the explanations are needed during the process, i.e., during design, installation or maintenance, and defence. Finally, the nature of the explanation is decided by answering the question of \textit{how} to interpret the AI/ML model. It will lay the groundwork for choosing the correct XAI methods for high-quality explanations. Our discussions later in this paper answer these questions for each aspect.

\subsection{Our Contribution}

To the best of the authors’ knowledge, this paper is the first of its kind to attempt to explore the capacity of XAI in a wide range of B5G security aspects. Table \ref{tab:surveys} depicts some of the relevant but dissociated studies carried out in this regard. However, none of them has been able to convey a holistic image of the role of XAI in B5G security. Therefore, our main contributions from this survey are listed below:

\begin{itemize}
    \item \textbf{Critically appraise the potential of XAI in the security domain}: This paper elaborates on the potential of XAI in the path to realizing accountability for AI/ML models are instrumental in enhancing network security and strengthening the resilience of 5G and future telecommunications. While numerous studies on 5G and beyond security incorporate data-driven ML solutions, there is limited emphasis on understanding the rationale behind their decisions. Serious doubts and questions regarding accountability arise with stakeholders when using black-box AI to secure 5G and beyond network components. We examine the ability of XAI methods to interpret black-box AI models (both pros and cons) in the context of 5G and beyond network security, addressing a significant and contemporary research gap. 

    \item \textbf{Comprehensively analyze XAI for commonly discussed 5G and beyond technical aspects}: Here, we explore the role of XAI in a range of B5G enabling technologies such as IoT/devices, \gls{RAN}, Edge network, core, and backhaul network, E2E slicing, and network automation. This list of enablers is carefully selected to cover most of the ground in 5G and beyond telecommunication architecture and provide a holistic view of the impact of XAI in 5G and beyond security. The study also incorporates the discussion with 5G and beyond use cases where necessary to provide a holistic comprehension to the reader. 
 
    \item \textbf{Survey of important, relevant research projects and standardizations}: Unlike in many other survey papers, here we explore the research projects that are underway to realize the 5G and beyond implementations and standardizations incorporating XAI. A detailed discussion of current projects and initiatives involving academic and industry partners provides clarity on the ongoing areas and the research gaps that are currently explored. AI security standardizations in 5G and beyond are discussed here to determine the requirements for future B5G networks and their respective technologies. 
    \item \textbf{Provide promising research directions as guidance}: Existing limitations and challenges with current XAI methods in security are exhaustively discussed, along with possible research directions. A few of the proposed research directions include security and isolation between network slices, computationally efficient explainable Edge-AI, and understanding the level of vulnerability of ML models to adversarial attacks in white-box and black-box contexts are some of the possible research directions that are identified. 
\end{itemize}

\begin{table*}[htbp]

\caption{Summary of important surveys on XAI for 5G and beyond Security}
\label{tab:surveys}
\renewcommand{\arraystretch}{1}
  \begin{tabular}{| p{0.4cm}|p{0.33cm}|p{0.33cm}|p{0.33cm}|p{0.33cm}|p{0.33cm}|p{0.33cm}|p{0.33cm}|p{10cm}|}
  \hline
      \rowcolor{Gray}
    	\multicolumn{1}{|c|}{\textbf{Ref.}} 
    	& \multicolumn{1}{|c|}{\textbf{Year}} 
         & \multicolumn{1}{|c|}{\textbf{{\rotatebox[origin=c]{90}{AI Techniques}}}}
         & \multicolumn{1}{|c|}{\textbf{{\rotatebox[origin=c]{90}{XAI Technical Aspects}}}}
         & \multicolumn{1}{|c|}{\textbf{{\rotatebox[origin=c]{90}{B5G Security Technical Aspects}}}}
         & \multicolumn{1}{|c|}{\textbf{{\rotatebox[origin=c]{90}{Role of XAI in B5G Security}}}}
         & \multicolumn{1}{|c|}{\textbf{{\rotatebox[origin=c]{90}{Challenges of XAI in B5G Security}}}}
         & \multicolumn{1}{|c|}{\textbf{{\rotatebox[origin=c]{90}{XAI/B5G Security Projects}}}}         
         & \multicolumn{1}{|c|}{\textbf{Remarks}}\\ [12ex]
    \hline
    \hline
    \multicolumn{1}{|c|}{\cite{guo2020explainable}} & 
        2020 &
        \cellcolor{yellow!30} M & 
        \cellcolor{green!30} H & 
        \cellcolor{red!30} L & 
        \cellcolor{red!30} L & 
        \cellcolor{yellow!30} M & 
        \cellcolor{red!30} L & 
        A review on motivation and framework for using XAI in 6G/wireless telecommunication for improving trust between humans and machines \\[2ex]
    \hline
        \multicolumn{1}{|c|}{\cite{das2020opportunities}} & 
        2020 &
        \cellcolor{yellow!30} M & 
        \cellcolor{green!30} H & 
        \cellcolor{red!30} L & 
        \cellcolor{red!30} L & 
        \cellcolor{yellow!30} M & 
        \cellcolor{red!30} L & 
        A comprehensible survey on various XAI techniques, their challenges and opportunities\\[2ex]
    \hline
        \multicolumn{1}{|c|}{\cite{de2021survey}} &
        2021 &
        \cellcolor{green!30} H & 
        \cellcolor{red!30} L & 
        \cellcolor{green!30} H & 
        \cellcolor{yellow!30} M & 
        \cellcolor{red!30} L & 
        \cellcolor{green!30} H & 
        A concise survey on the 6G future trends, applications, requirements and technical aspects.\\[2ex]
    \hline
        \multicolumn{1}{|c|}{\cite{nguyen2021security}} &
        2021 &
        \cellcolor{yellow!30} M & 
        \cellcolor{red!30} L & 
        \cellcolor{green!30} H & 
        \cellcolor{green!30} H & 
        \cellcolor{red!30} L & 
        \cellcolor{red!30} L & 
        A survey on technologies and challenges in 6G security and privacy in different layers of 6G architecture\\[2ex]
    \hline
    \multicolumn{1}{|c|}{\cite{arrieta2020explainable}} &
        2022 &
        \cellcolor{green!30} H & 
        \cellcolor{green!30} H & 
        \cellcolor{red!30} M & 
        \cellcolor{yellow!30} M & 
        \cellcolor{yellow!30} M & 
        \cellcolor{red!30} L & 
        A survey on explainable AI over the Internet of Things (IoT): Overview, state-of-the-art and future directions\\[2ex]
    \hline
    \multicolumn{1}{|c|}{\cite{fiandrino2022toward}} & 
        2022 &
        \cellcolor{yellow!30} M & 
        \cellcolor{green!30} H & 
        \cellcolor{yellow!30} M & 
        \cellcolor{red!30} L & 
        \cellcolor{red!30} M & 
        \cellcolor{yellow!30} L & 
        A survey on explainable and robust AI in 6G networks discussing the current state, challenges and future work\\[2ex]
    \hline
    \multicolumn{1}{|c|}{\cite{garg2023trusted}} &
        2023 &
        \cellcolor{green!30} H & 
        \cellcolor{yellow!30} M & 
        \cellcolor{red!30} M & 
        \cellcolor{red!30} L & 
        \cellcolor{red!30} L & 
        \cellcolor{red!30} L & 
        A review on trusted Explainable AI for 6G-Enabled Edge Cloud Ecosystem \\[2ex]
    \hline
        \multicolumn{1}{|c|}{\cite{shahriar2023survey}} & 
        2023 &
        \cellcolor{green!30} H & 
        \cellcolor{red!30} M & 
        \cellcolor{green!30} H & 
        \cellcolor{yellow!30} M & 
        \cellcolor{red!30} L & 
        \cellcolor{green!30} L & 
        A survey of privacy risks and mitigation strategies in the Artificial intelligence life cycle\\[2ex]
    \hline
    \multicolumn{1}{|c|}{\cite{khan2023explainable}} & 
        2024 &
        \cellcolor{green!30} H & 
        \cellcolor{yellow!30} M & 
        \cellcolor{red!30} L & 
        \cellcolor{red!30} L & 
        \cellcolor{red!30} L & 
        \cellcolor{red!30} L & 
        Insightful discussion on explainable and Robust Artificial Intelligence for Trustworthy Resource Management in 6G Networks\\[2ex]
    \hline
    \multicolumn{1}{|c|}{\cite{wang2024explainable}} & 
        2024 &
        \cellcolor{green!30} H & 
        \cellcolor{green!30} H & 
        \cellcolor{yellow!30} M & 
        \cellcolor{red!30} L & 
        \cellcolor{red!30} L & 
        \cellcolor{green!30} H & 
        A survey on explainable AI for 6G use cases, technical aspects and research challenges\\[2ex]
    \hline
        \multicolumn{1}{|c|}{\textbf{This paper}} & 
        2024 &
        \cellcolor{green!30} \textbf{H} & 
        \cellcolor{green!30} \textbf{H} & 
        \cellcolor{green!30} \textbf{H} & 
        \cellcolor{green!30} \textbf{H} &
        \cellcolor{green!30} \textbf{H} &
        \cellcolor{green!30} \textbf{H} &
        \textbf{A comprehensive survey of using XAI for trustworthy and transparent 6G security including use cases, requirements/vision, technical aspects, projects, research work, standardization approaches and future research directions.}\\[2ex]
    \hline
\end{tabular}

\begin{flushleft}
\begin{center}
    
\begin{tikzpicture}

\node (rect) at (1,2) [draw,thick,minimum width=1cm,minimum height=0.7cm, fill= red!30, label=0:Low/No Coverage] {L};
\node (rect) at (6.3,2) [draw,thick,minimum width=1cm,minimum height=0.7cm, fill= yellow!30, label=0:Medium Coverage] {M};
\node (rect) at (12,2) [draw,thick,minimum width=1cm,minimum height=0.7cm, fill= green!30, label=0:High Coverage] {H};
\end{tikzpicture}
\end{center}

\end{flushleft}
  
\end{table*}

\subsection{Paper Outline}

This section introduces the motivation and contribution of this survey paper. The second section provides background on the technical aspects of this paper, namely, B5G, XAI, and XAI's potential for improving B5G security. Then, the details of these technical aspects are discussed in sections \ref{sec:xai_role_network} to \ref{sec:xai_role_other}.
Sections from \ref{sec:xai_role_network} to \ref{sec:xai_role_other} analyze the impact of introducing XAI on the existing AI-powered security solutions in the network layer, cross layers and traditional security aspects of 5G and beyond networking paradigm. Section \ref{sec:new_issues} highlights potential new security issues because of introducing XAI. 
Section \ref{sec:projects} strengthens the importance of this survey paper by listing the ongoing research projects and standardizations about 5G and beyond security and XAI. Section \ref{sec:lesson} summarises sections from \ref{sec:xai_role_network} to \ref{sec:new_issues}, and \ref{sec:projects} with the lessons learned and future research directions. Finally, section \ref{sec:conclusion} concludes the whole survey.

 \begin{figure}[htb]
    \centering
    \includegraphics[width=0.48\textwidth]{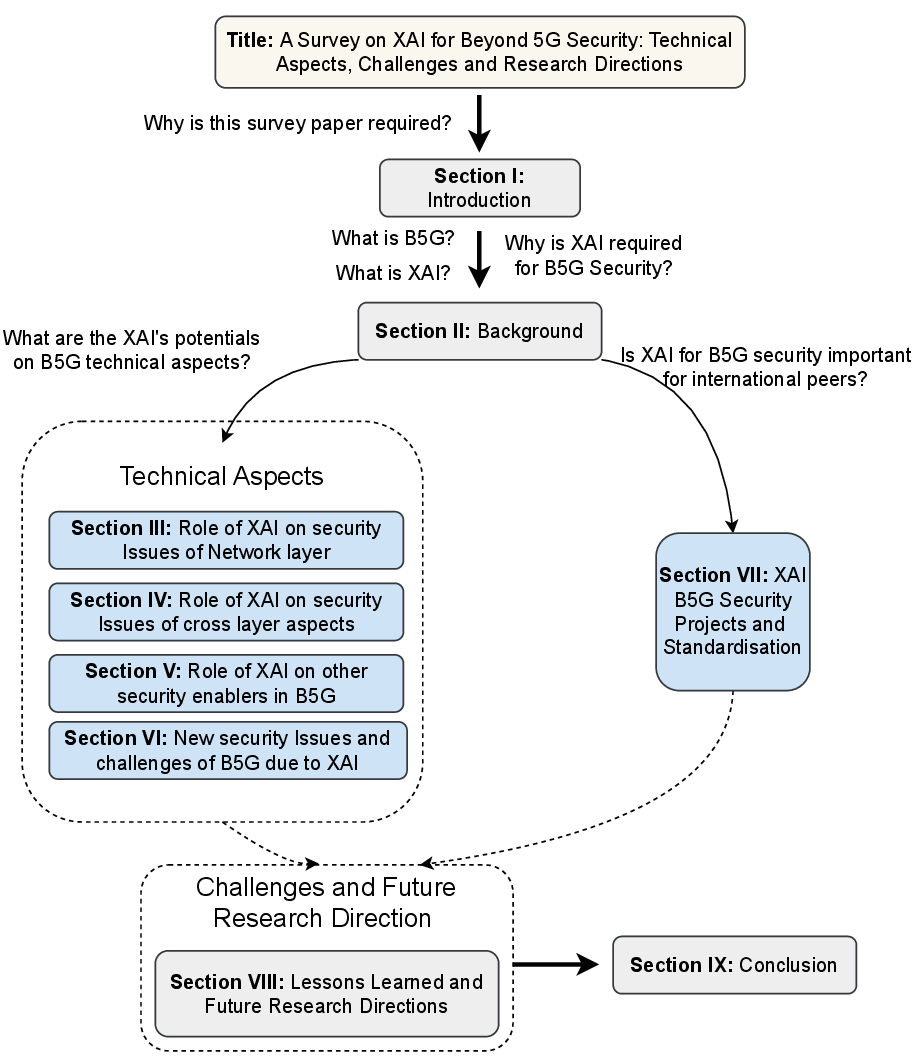}
    \vspace{1mm}
    \caption{This figure outlines the paper structure. We lay down the context with motivation for XAI in the B5G security, our contributions, and the outline for the paper. Stemming from the theme set in the introduction, we answer the questions of whats and whys for XAI in the background section. The rest of the paper extends the minutiae of the XAI's potential in B5G security aspects, current standardizations, and projects. Finally, the Lessons learned and future research directions conclude the main takeaways of the survey. }
    \label{fig:paper_outline}
\end{figure}

\section{Background} \label{sec:bg}

This section briefly introduces the background of the related technologies discussed in this paper. In particular, 5G and beyond technologies and XAI concepts are discussed, followed by a discussion on the growing need for XAI for 5G and beyond security.

\subsection{5G and beyond}

The rapid growth of the communication industry in the last decade has enabled 5G technologies to be widely commercialized in recent days. Following the success of 5G, 6G/B5G is becoming the focal point of academia and industry with research and implementations. 5G has addressed much of the prevalent problems \cite{ho2019next} with high data rate enhanced mobile broadband systems (eMBB) and leapt on with new functionalities such as laying the foundation for enabling the Internet of Things (\gls{IoT}). New \gls{IoT} services are developed rapidly in applications such as virtual, augmented, mixed reality services (which fall under \gls{XR} services), autonomous vehicle systems, brain-computer interfaces (BCI), telemedicine, haptic systems and blockchain-based systems \cite{saad2019vision}. In order to implement these services,  ultra-reliable, low-latency communications (URLLC) with short-packet support and high data rates in both uplink and downlink need to be maintained in a secure and privacy-protected wireless system \cite{ho2019next}. The massive number of human and machine-type devices connecting to the network will shape the revolution. After the full deployment of 5G, URLLC and Massive Machine Type Communication (mMTC) will address those devices' end-to-end latency needs. That means in the real world beyond 5G, for example, in 6G technologies, data rates must reach terabytes (maximum 1 Terabit/second) to effectively serve heterogeneous devices. In other words, nearly a 1000x increase from the last generation of wireless technologies \cite{chowdhury20206g} bringing in massive amounts of data each day. A cohort of technologies like AI, Symbiotic radio (SR), call-free massive MIMO (CFmMM), intelligent communication surfaces, index modulation (IM), simultaneous wireless information and power transfer (SWIPT), network-in-box \cite{dang2020should, shafin2020artificial, de2021survey, kato2020ten, tan2020thz, gui20206g, chen2020vision, you2021towards} will be used in handling those services mentioned above. AI takes a prominent place out of them due to its proven unprecedented capabilities.

Following the massive success of AI in computer vision, natural language processing, speech recognition, bioinformatics, social intelligence, and numerous others, the technology has proved to be ubiquitous \cite{zhang2020artificial}. Due to the vast and varied set of applications associated with billions of devices in the 5G and beyond eco-system, a tremendous amount of data will be generated at high rates, making it ideal for AI for AI-based problem-solving. 

\subsection{Explainable AI}
\subsubsection{Motivations of XAI}

While the early AI systems were simple to understand, opaque decision methods such as \gls{DNN} have recently gained popularity. \gls{DL} models are experimentally successful due to a combination of efficient learning algorithms and their large parametric field.
DNNs are considered sophisticated black-box models since they have hundreds of layers and millions of parameters \cite{castelvecchi2016can}. Transparency is the polar opposite of black-boxness, which is the pursuit of knowledge of how a model functions. The need for explainability among AI stakeholders is growing as black-box \gls{ML} algorithms are increasingly used to make significant predictions in critical settings \cite{preece2018stakeholders}. The risk lies in making and implementing choices that are not reasonable, lawful, or do not allow for comprehensive explanations of their actions \cite{gunning2019darpa}. Explanations that back up a model's output are critical. For example, in medical applications, specialists need to uncover the causes in the model to arrive at the forecast, reinforcing their confidence in the diagnosis \cite{tjoa2020survey}. Telecommunication systems, B5G-backed autonomous cars, security, and finance are just a few other examples. 

Interpretability in machine learning model implementation enhances model debugging by providing insights for decision impartiality, correcting training dataset bias, and generalizing ML solutions. XAI outputs ensure relevant variables are used, model reasoning is causal, and can detect adversarial events in network and security domains\cite{arrieta2020explainable, guidotti2018survey}. XAI outputs improve clients' trust in models, and high-impact stakeholders benefit from XAI's role in security audits and regulatory processes, enhancing fairness and ethics during model development and data collection during various parts of wireless data-driven life cycle

Design-interpretable models are distinguishable from externally explicable ones. Other categorizations are listed below. Each category has beneficial characteristics in different situations. Proprietary model owners may not want to share model architectures, limiting the XAI methods to post-hoc XAI methods. A more comprehensive array of XAI methods (in addition to post-hoc XAI methods) may be used for open-source models.

\subsubsection{Transparency}

Rule-based models and transparent models are two types of models that provide interpretability and expressiveness. Rule-based models can be understood independently and are categorized into decomposable, simulatable, and algorithmically transparent models. Decomposable models can be explained in terms of their constituent components, simulatable models can be simulated or thought about rigorously, and algorithmically transparent models are entirely explicable using mathematical methods. Popular transparent models include Linear/Logistic regression, Decision Trees, K-nearest neighbours, Rule-based models, GAM, and Bayesian models.

\begin{figure}[htb]
    \centering
    \includegraphics[width=0.45\textwidth]{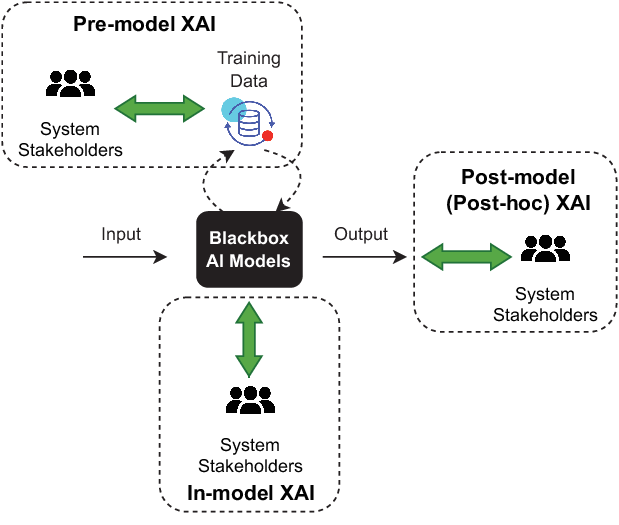} 
    \vspace{1mm}
    \caption{XAI Taxonomy. Pre-model XAI explains the training data used to build AI models (e.g.,Principal component analysis (PCA), and t-Distributed Stochastic Neighbor Embedding (t-SNE) ). In-model XAI refers to transparent AI models that are self-explanatory (e.g., decision trees, random forests). Post-hoc XAI models explain the results of the trained AI models (e.g., LIME, SHAP).}
    \label{fig:xai_tax}
\end{figure}

\subsubsection{Taxonomy of XAI}
The XAI methods can be divided into multiple categories based on various criteria \cite{arrieta2020explainable, singh2020explainable }. The most common XAI-based taxonomies are discussed below. XAI methods that fall into those categories are not necessarily exclusive for each group. According to the taxonomy, some methods can belong to even two or more categories. 

\begin{enumerate}[(a)] 

\item \textit{Model-agnostic vs Model-specific:}
Model agnostic methods for XAI, being flexible, are adept at decoding black box models' decision processes regardless of the model type. On the other hand, model-specific methods, tailored to specific models, use core components to interpret outcomes, making them ideal for identifying granular aspects. However, this specialization comes at the cost of flexibility.

\item \textit{Local vs. Global methods:}
Local methods, which interpret specific data points, are designed to explore the ML function's close proximity. They are faster but can be erratic. In contrast, global methods consider the entire ML function, making them slower but more robust in their interpretation.

\item \textit{Pre-model, In-model vs Post-model explainers:}
XAI methods can be applied at three stages of the AI lifecycle: pre-model, in-model, and post-model. Pre-model methods aid in data analysis and feature engineering before the models are trained, while in-model methods are embedded in ML algorithms or intrinsically explainable, like linear regression and decision trees. Post-hoc/post-model explanations interpret models after they are trained most of the time using query-level access.

\item \textit{Surrogate vs Visualization:}
\label{sec:XAI_taxonomy_surrogate_viz}
The division between surrogate and visualization is based on how explanations are generated. Surrogate model-based explainers generate explanations from an approximated model of the original model, and visualization techniques, which explore the model's internal workings, use the original model and data to interpret.

\end{enumerate}
\subsubsection{XAI Methods}
Numerous methods are studied in the literature to explain black-box AI/ML models. Here we have summarised a selected set of popular XAI methods for supervised and reinforced learning AI that are more established in the academic and industrial community, as shown in Table. \ref{tab:xaiMethods}. We have also discussed XAI for unsupervised learning AI seperately. 

\newcolumntype{s}{>{\hsize=.4\hsize}X}

\begin{table}[!ht]

\caption{Summary of most popular XAI methods}
\label{tab:xaiMethods}
\begin{tabularx}{0.45\textwidth}{sX}
	\midrule
	\textbf{Layer} & \textbf{Important Characteristics} \\
	\midrule
    LIME - Locally Interpretable Model Agnostic Explanations \cite{ribeiro2016should}& \begin{itemize}[noitemsep,nolistsep] \item{}Model Agnostic
    \item{}Post-hoc explanation
    \item{}Can be used to explain supervised AI/Ml algoritms
    \item{}Local interpretation only
    \item{}Uses Surrogate model
    \item{}Efficient calculation speed
    \end{itemize} \\
	\midrule
    SHAP - Shapley Additive Explanations \cite{ lundberg2017unified }& \begin{itemize}[noitemsep,nolistsep]
      \item{}Model Agnostic 
      \item{}Post-hoc explanation
      \item{}Can be used to explain supervised AI/Ml algoritms
      \item{}Local/Global Interpretability
      \item{}Uses Surrogate model
      \item{}Computationally expensive
      \end{itemize} \\
	\midrule
    LRP - Layer-wise Relevance propagation  \cite{ binder2016layer }& \begin{itemize}[noitemsep,nolistsep]
      \item{}Model specific
      \item{}Post-hoc explanation
      \item{}Can be used to explain DL algorithms
      \item{}Local Interpretability
      \item{}Visual Explanations
      \end{itemize} \\
      	\midrule
    CFE - Counterfactual Explanations & \begin{itemize}[noitemsep,nolistsep]
      \item{}Both model specific and agnostic methods are available
      \item{}Post-hoc explanation
      \item{}Can be used to explain DL algorithms
      \item{}Local/Global interpretability
      \item{}Uses Surrogate model
      \end{itemize} \\
      	\midrule
    XAI FOR RL - Various methods of XAI for Reinforcement learning& \begin{itemize}[noitemsep,nolistsep]    
      \item{}PIRL (Programmatically Interpretable Reinforcement Learning) \cite{verma2018programmatically}
      \begin{itemize}[noitemsep,nolistsep]
        \item{Model specific}
        \item{in-model}
        \item{global}
        \item{surrogate explainer}
      \end{itemize}
      \item{}Heirrachial Policies Technique \cite{ shu2017hierarchical }
      \begin{itemize}[noitemsep,nolistsep]
        \item{Model specific}
        \item{in-model}
        \item{local}
        \item{surrogate explainer}
      \end{itemize}
      \item{}LMUTs - Linear Model U-Trees
      \begin{itemize}[noitemsep,nolistsep]
        \item{Flexible}
        \item{post-hoc}
        \item{local/global}
        \item{surrogate explainer}
      \end{itemize}
      \end{itemize} \\
	\bottomrule
\end{tabularx}

\end{table}

\textit{XAI for Unsupervised learning}
XAI for unsupervised learning is still in its infancy. Unsupervised learning techniques such as clustering help understand unlabeled data clearly. However, one can argue that the need for explainability is even higher for unsupervised learning since they are tough to validate quantitatively \cite{von2012clustering}. Pre-model techniques like \gls{PCA} help to visualize clusters in lower dimensions than the original data dimensionality; however, drawbacks such as information losses, missing non-linear relationships, and effect of outliers exist in PCA as an XAI method \cite{rao1964use, ding2004k}. Tree-based clustering methods proposed in \cite{frost2020exkmc, dasgupta2020explainable, loyola2020explainable} can be considered in-model explanation methods. They can also interpret complex models such as unsupervised \gls{VAE} \cite{nguyen2019gee, curi2019interpretable, neumeier2021variational}. There's evidence that explanations can be incomplete in some cases \cite{neumeier2021variational}. Post-model explainers usually follow the three-step approach of clustering data, training a classifier using output cluster names as labels, and generating post-hoc explanations using any post-hoc XAI method. This method has been tested with various model types and XAI methods, including LRP, LIME/SHAP \cite{montavon2022explaining, morichetta2019explain, horel2020explainable}, but is still in its infancy and may suffer from biases.

In conclusion, XAI is a field that is consistently expanding. We have summarized some of the emerging XAI methods in Table.~\ref{tab:xai_methods_tab} that can be considered relevant to the security of the 5G and beyond. Most listed techniques are specific to time series and tabular data types, often seen in the networking domain. However, the currently available XAI methods are primarily suitable for low-level users directly. Some projects working towards making explanations more comprehensive to end-users are discussed at the end of this paper.

\newcolumntype{b}{X}
\newcolumntype{s}{>{\hsize=.1\hsize}X}
\newcolumntype{m}{>{\hsize=.25\hsize}X}
\newcolumntype{n}{>{\hsize=.3\hsize}X}
\newcolumntype{d}{>{\hsize=.25\hsize}X}
\renewcommand\theadalign{t}

\begin{table*}[]

\caption{XAI methods, their characteristics and potential applications in security of 5G and beyond networks}
\label{tab:xai_methods_tab}
\begin{tabularx}{0.98\textwidth}{mnsdlX}
\toprule
\textbf{Method} & \textbf{Explanation type} & \textbf{\thead{MA/\\ MS}} & \textbf{Direct user level} & \textbf{Data types} & \textbf{Potential Security Applications} \\ \midrule
SHAP\cite{lundberg2017unified} & Perturbations-based, Local/global & MA & Security system developer/ operators/ & \begin{tabular}[c]{@{}l@{}}Tabular,\\ Image, text,\\Genomic\end{tabular}  & Malware/virus spam detection in ORAN, NFV, VNFattacks detection in 5G and beyond, Malicious Microservice detection, Data breach detection, Suitable mostly cloud based operations that have sufficient compute power \\
LIME \cite{ribeiro2016should}& Perturbations-based, Local & MA & Security system developer/operators & \begin{tabular}[c]{@{}l@{}} Tabular, \\Image, text\end{tabular}& Edge security,   Space-air communication, IDS/MTD \\
CoMTE \cite{ates2021counterfactual} & Counterfactual, Local & MA & Security system developers & Time-series & Proactive intrusion prevention, ORAN/cloud malicious policy detection, Encrypted traffic inspection in ORAN and cloud\\
C-CHVAE \cite{pawelczyk2020learning} & Counterfactual, Local & MA & Security system developers & Tabular & Core and backhaul network NIDS protection, protocol vulnerability detection in , Deep packet inspection \\
PDL \cite{ gee2019explaining} & Prototype, Global, In-model & MS & Security system developers /operators & Time-series & Trusted update and program verification in AI driven networks,  Risk-based authentication\\
AFEX \cite{ konstantinov2023attention} & Attention mechanism, GAM, Local/Global & MS & Security system   developers & Tabular data & 5G and beyond proactive intrusion prevention, biometric authentication, network slice anti-virus/malware/malicious content detection \\
CA-SFCN \cite{ hao2020new}& Attention Mechanism, Local & MS & Security system developers & Time-series &  5G and beyond network Malware/spam detection, Runtime protection, mmWave Beamforming alignment \\
FAHP \cite{ el2018ontology } & Fuzzy logic, Global & MS & end users & Time-series & IDS, Protocol   vulnerability detection, Encrypted traffic prediction, signal detection in physical layer security \\
Deep FCM \cite{wang2020deep} & Fuzzy logic, Global & MS & Security system developers & Time-series & ORAN XApp specialized IDS, Deep packet inspection, Misbehaviour detection \\
BB-BC IT2FLS \cite{ ferreyra2019depicting} & Fuzzy logic, Global & MS & end users & Tabular & Anti-jamming, physical layer authentication, Protocol vulnerability detection \\
SAX-VSM \cite{senin2013sax} & SAX, In-Model, Global & MS & Security system developers & Time-series & Future network edge security, Container protection in NFV  \\
Tsxplain \cite{ munir2019tsxplain } & Backpropagation-based, Local & MA & Security system developers/operators & Time-series & Deep packet inspection for ORAN, Encrypted traffic inspection, Beamforming alignment in 5G and beyond \\
FCN \cite{ismail2019accurate} & Backpropagation-based, Local & MS & Security system   developers & Time-series & Signal detection in mmWave beamforming and RIS in 6G, Misbehaviour analysis on , traffic analysis \\ \bottomrule

\end{tabularx}

\begin{minipage}{12cm}
\vspace{0.1cm}
\small MA - Model Agnostic, MS - Model Specific
\end{minipage}

\end{table*}


\subsubsection{Stakeholders of XAI}
The research communities are actively working on explainable AI, with various stakeholders enhancing, evaluating, regulating, and manipulating AI in various applications. The level of explainability and interpretability is influenced by various stakeholders. It is crucial to identify the parties involved in the full wireless data lifecycle of 5G and beyond for security to improve accountability and trustworthiness. Here, we define five main stakeholder communities: system creators, system operators, theorists, ethicists, and end-users. \cite{preece2018stakeholders, tomsett2018interpretable }.

\begin{figure}[htb]
    \centering
    \includegraphics[width=0.49\textwidth]{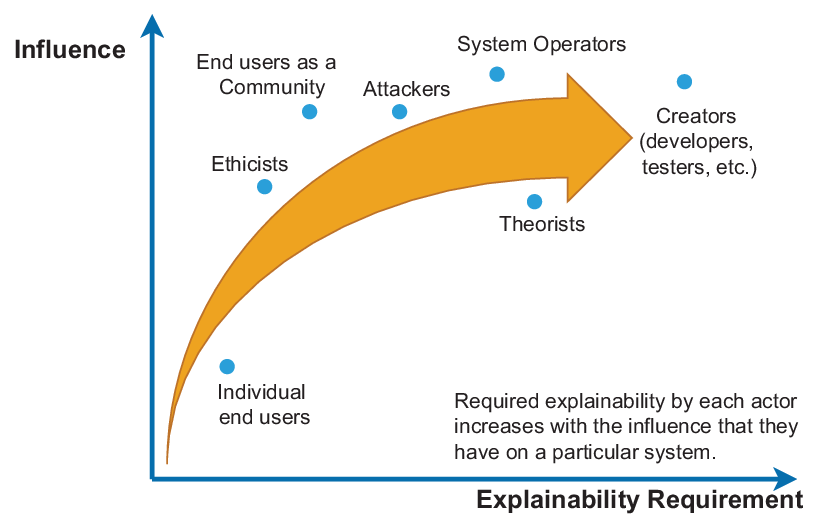} 
    \vspace{1mm}
    \caption{XAI Stakeholders: Different levels of influence that each stakeholder has on the systems and their respective explainability requirements. 
    }
    \label{fig:stakeholders}
\end{figure}

\begin{itemize}

\item \textit{Creators:}. Creators are those who build secure, high-fidelity AI-based 5G and beyond applications. This group includes implementers (developers, testers, security experts, data scientists) and owners (agents, business owners) working on AI/ML applications. Creators can have roots in the industry or academia. Their impartiality and resilience requirements are of the highest regard. Their influence on XAI is also very high. 

\item \textit{System Operators:} System operators maintain the systems and ensure smooth operation after deploying an AI/ML-based system. Although they might not require a granularity of explanations as high as developers, they still require a high enough explainability to detect and verify anomalies in the system to provide runtime solutions. Similarly, the influence on the system and data can be considered moderately high.

\item \textit{Theorists:} Theorists are those who are interested in comprehending and expanding AI theory, especially as it relates to \gls{DNN}s. Members of this group are often associated with university or industry research institutions. This community requires a high level of explainability. Their influence on the XAI, in general, can be considered to be high.

\begin{figure*}[b]
    \centering
    \includegraphics[width=0.98\textwidth]{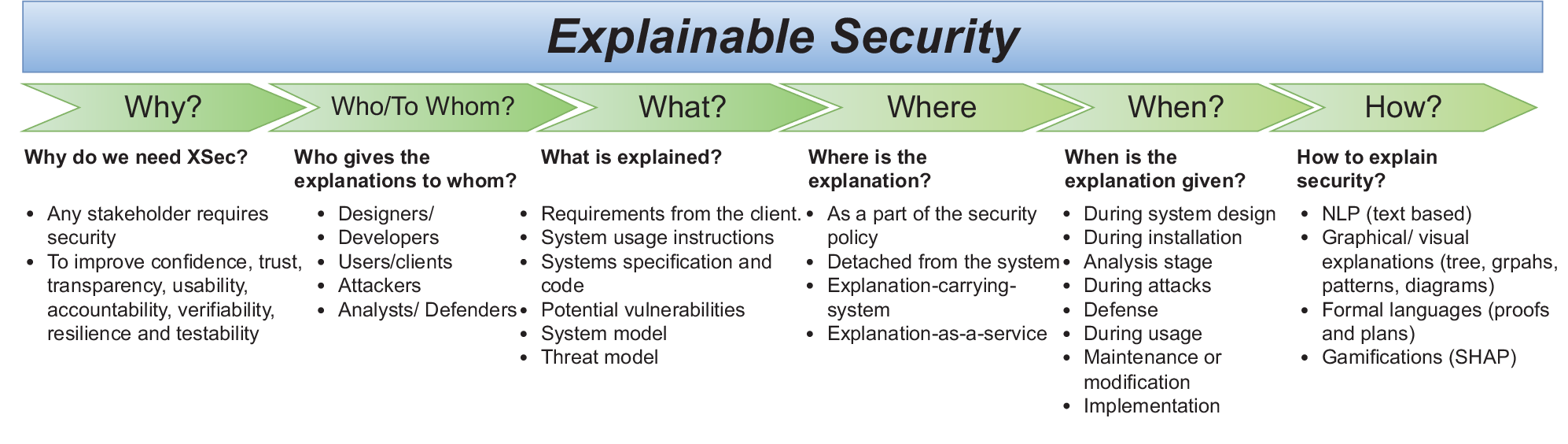}
    \vspace{1mm}
    \caption{6W analysis for explainable security in 5G and beyond. The procedure shown can be used as a framework to initiate laying the groundwork when designing security aspects of explainable intelligent systems built in/on 5G and beyond network
    }
    \label{fig:xsec}
\end{figure*}

\item \textit{Ethicists:} Ethicist observers advise, comment, and criticize AI systems on fairness, accountability, and transparency. This group includes computer engineers, scientists, social scientists, attorneys, journalists, economists, and politicians with ethical concerns about AI models. To ensure fairness, impartiality, and comprehensible disclosure for accountability and auditability, ethicists must explain beyond software quality. This stakeholder group also includes organizations like the EU's GDPR \cite{goodman2017european} or the US DARPA Regulations \cite{gunning2019darpa}. Their influence on AI systems is significant. 

\item \textit{End Users:} Finally, the users need explanations to assist them in deciding whether or how to act in response to the systems' outputs and to assist them in justifying their actions. This community comprises everyone engaged in processes affected by an artificial intelligence system. The explainability requirement for end-users is similar to that of ethicists; however, their influence on the system is only strong under particular circumstances (e.g., community/group approach). 

\end{itemize}

In light of the preceding discussion, the most logical approach may be to provide different explanations tailored to the various stakeholders. Nevertheless, it is also possible to envision a composite explanation object containing all of the required information to satisfy multiple stakeholders at once.

\subsection{Tutorial (with code) for getting started with XAI for security}

We have implemented a practical example (refer to footnote 1) of using XAI methods as a tutorial to understand better the influence of specific features on the inferences a neural network makes for network intrusion detection. This Tutorial will be a stepping stone for implementing much more sophisticated use cases of rather elaborate security functions such as malicious action detection, defence, and reconciliation. Here, we build a neural network trained on the NSL-KDD dataset \cite{NSL2009}. Then we interpret it to create SHAP \cite{lundberg2017unified} and LIME \cite{ribeiro2016should} explanations. This Tutorial acts as a foundation for anyone who wants to explore using XAI for security in the B5G. The code relevant to the Tutorial is found in GitHub\footnote[1]{\label{tutoriallink}Link to the tutorial code: https://github.com/t-T-s/xai\_tutorial} repository. Using explanations, we show that an operator can monitor the ML model's health and reveal vital attribution-based information to the user. 
\section{Role of XAI on Security Issues of Network Layer} \label{sec:xai_role_network}

\subsection{Security of Radio Access Network}

5G and beyond future networks leverage open standards, virtualization, and AI/ML to create a more flexible, interoperable, and intelligent network infrastructure and \gls{RAN} spearheads this innovation. \gls{RAN} comprises components of a telecommunications system that link mobile devices/\gls{UE} to public and a private core network via an existing network backbone. LTE and 5G RANs can offer ultra-reliable (deterministic) wireless performance \cite{sdxcentral2021}. Subsequently, many kinds of RANs, including Enhanced Data Rates for GSM Evolution RAN (GERAN), Universal Mobile Telecommunications System RAN (UTRAN), and Evolved UTRAN (E-UTRAN) have been implemented as 2G, 3G, and 4G radio access technologies have progressed. The latest additions include \gls{CRAN}, \gls{VRAN}, and \gls{ORAN}, which are anticipated to be linked to 5G and future generations of wireless technology \cite{celona2021}. Our analysis concentrates explicitly on the ORAN advancements.

\subsubsection{Possible Security Threats, Challenges, Issues}

A RAN can consist of a base-band unit (BBU), radio unit, or remote radio unit, antennas, and software interfaces. One of the earliest RANs was the Global System for Mobile Communications (GSM) RAN. 

RAN protocol stack is proposed to split into three functional blocks: CU (Central Unit), DU (Distributed Unit), and RU (Radio Unit) to allow increased flexibility and scalability in combining software and hardware from different vendors. With its \gls{RIC}, open interfaces, and disaggregated design, O-RAN ultimately allows the realistic deployment and execution of AI/ML solutions at scale. These solutions either infer and anticipate network traffic or dynamically change the nodes of RAN depending on real-time settings and user demands. 

\begin{figure*}[htb]
    \centering
    \includegraphics[width=0.99\textwidth]{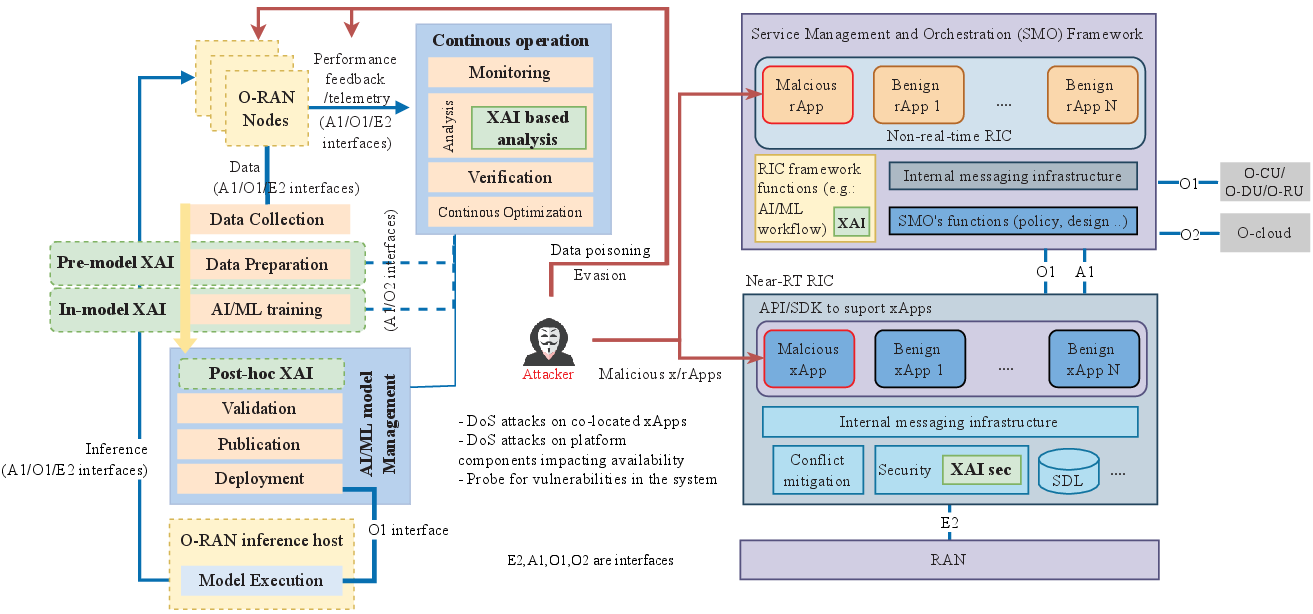}
    \vspace{1mm}
    \caption {Depiction of the ORAN architecture with an additional layer of XAI-based security. The left-hand block diagram shows the AI/ML workflow in ORAN, and the right-hand figure shows the overall architecture for ORAN. Input data obtained through the RAN nodes via the E2 interface are vulnerable to poisoning and manipulation by adversaries. With pre-model XAI, some of the problematic data can be filtered out. Furthermore, any undetected threats can be detected through post-hoc/in-model XAI methods (in Non-RT RIC) during the validation phase before deploying them to the inference host. Verification of x/rApps that use black-box (for either proprietary reasons or complexity) models can be aided with the XAI methods that unveil the reasons for the model's adversarial behaviours. This would also avoid subsequent attacks, such as DoS attacks, which can cause control and policy conflicts in RICs.\cite{oranalliance2021wg2AIML, polese2022understanding}}
    \label{fig:ORANXAI}
\end{figure*}

We focus on the threats and challenges in CRAN, ORAN, and VRAN. The C-RAN architecture can be affected by various security threats \cite{tian2017survey}. Some examples are eavesdropping, \gls{MITM} attacks, DoS, MAC spoofing, identity theft attacks, jamming attacks, and TCP/UDP flooding. However, some threats are inherited from the predecessors of \gls{CRAN} and Cognitive Radio Networks (CRN). The ORAN specification defines a new class of threats: attacks on AI/ML models utilized for inference and control in xApps and rApps in ORAN. Poisoning attacks are one of the primary attacks discussed here. In this attack, an adversary takes advantage of unregulated access to the data stored in the service and management orchestration or non-real-time \gls{RIC} to inject altered and misleading data into the datasets used for offline training of AI/ML algorithms. An attacker might also take control of one or more O-RAN nodes (\ref{fig:ORANXAI}) to produce synthetic data for online AI/ML fine-tuning or inference. These attacks may cause AI and ML systems to generate inaccurate predictions or control decisions, leading to performance issues or outages \cite{polese2022understanding}. Data from various ORAN split components (RU, CU, DU) are used for multiple inference functions. For example, compressing I/Q signals in the front haul can cause significant risks for RAN intelligence in reducing the impact from noise \cite{rodriguez2020cloud} that is generated by a jamming attack. In such cases, more insights can be drawn by using XAI, such as which features, out of the most essential features for the AI/ML model, are affected by the noise.

\gls{RIS} is an emerging technology that manipulates and reflects wireless signals to enhance coverage, capacity, and energy efficiency, enhancing the 5G enablers such as mmWave communication in the physical aspects of RAN \cite{tang2021novel}. However, the intricacies involved in optimizing the RIS can be aided with XAI, particularly when considering a scenario where multi-user multi-RIS with non-deterministic receivers in dynamic environments are involved. RIS holds the potential to significantly improve signal strength, extend coverage, and enhance the overall spectral efficiency of 5G and beyond networks, especially with mmWave communication \cite{elmossallamy2020reconfigurable}. Current ML-based beamforming methods are prone to biases and attacks that exploit them \cite{kuzlu2023adversarial}. Work such as \cite{catak2022security, kim2021adversarial, dinh2023defensive, hoang2022detection} have shown that beam training in RIS mmWave networks is vulnerable to different attacks, including jamming and spoofing. For example, in \cite{kim2021adversarial}, a deep neural network was trained to improve beam selection robustness and latency. They show that a well-crafted perturbation that an attacker adds to the data can prevent the genuine 5G user from correctly classifying beam patterns. LIME can monitor the input data and maintain a stable user connection under these malicious circumstances.

Machine Learning-based IDS is the most promising anomaly-based IDS because it can gradually improve its performance by learning over time while performing a given task. Authors of \cite{hachimi2020multi} have used \gls{MLP} and \gls{SVM} enabled with kernel trick (KSVM) to classify and detect multi-stage jamming attacks in CRAN BBU pool. Regarding O-RANs, self-organization and intelligence-based technologies will be extensively used in the deployment process \cite{gavrilovska2020cloud}. 

\subsubsection{How XAI can help to mitigate these attacks/issues}

The AI/ML workflow standardized by ORAN WG2 (Working Group 2) \cite{oranalliance2021wg2AIML} shines a light on the importance of Trusting AI for the evolution of RAN. From the ORAN architecture in figure \ref{fig:ORANXAI}, it becomes clear that AI/ML are primarily deployed in the \gls{RIC} as xApps/rApps or as dApps \cite{d2022dapps} directly in the CUs/DUs. x/rApps will eventually bring the vision of self-organizing networks to fruition. These networks can autonomously detect ongoing changes in the state of the channel, the network, and the traffic. 

Among others, functions include handover and mobility management, spectrum coexistence, network slicing, and resource allocation. Surgically manipulating the AI/ML models deployed in the RIC by adversaries could disrupt the node bases, resulting in network failures. It is more of a question of \textit{when} it will happen, and that would be where XAI comes into play. With XAI methods such as LIME, SHAP, and counterfactual explanations, the stakeholders can identify the problems with black-box AI/ML solutions using different \gls{KPM} in higher granularity. For example, channel quality information, modulation and coding schemes, throughput, latency, data demand, and jitter are a few of the KPMs \cite{polese2022understanding} whose behaviours as features can be uncovered through model gradients, training samples, or attribution methods in opaque models through XAI. Not just the feature contributions but also the bounds for a \gls{KPM} to flip decisions. This output alerts security operators when the values suspiciously shift toward the boundaries due to poisoning or adversarial attempts. 
Some attacks, such as jamming, spoofing, and DoS attacks \cite{oranalliance2021wg11secspec}, were identified through AI and ML-based IDSs. However, in the wireless AI-drivenAI-driven environment, these IDSs are also susceptible to adversarial attacks such as backdoors. An adversarial input to exploit the poisoned backdoor of an ML model can be traced back to the original training samples through sample importance-based XAI methods.

This type of root cause analysis provides the solution to the problem \textit{What would happen when the system fails?}. To summarise, XAI provides essential knowledge on security parameters, such as the detectability of attacks through transparency and accountability, as well as availability and integrity for various instances. For example, XAI will be integral in monitoring the availability of infrastructure during malicious deployment of black-box x/rApp. Any intentional conflict between control sequences in x/rApps and unauthorized access to disaggregated RAN components are solvable with more explainable systems.

Although such attacks can be detected and defended successfully \cite{dinh2023defensive }, the question remains whether they can always be considered accountable and trustworthy. These methods use more ML models in the defensive actions. For example, the receivers (Automated Guided Vehicles) blindly trust the outputs of the two AEs used in \cite{dinh2023defensive } for detection and defence against jamming attacks, disregarding the fact that autoencoders themselves are vulnerable to attacks/biases. For instance, the effect on received signal strength (RSS) can unexpectedly fluctuate based on multiple feature values not seen during training, making the root cause of any misclassification practically impossible. During deployment, the sheer complexity of AEs can make the debugging process tedious. During actual world implementations, ground truths are generally unavailable. Thus, the feature contributions obtained through XAI would shine a light on the black-box AEs under such conditions, providing clues to recognize which features have been tampered with.

\subsubsection{Added cost of Using XAI}

AI/ML-powered radio resource allocation, resource scheduling, and power allocation are integral functions of ORAN (Fig. \ref{fig:ORANXAI}). To ensure accountability, open distributed units hosting those models will also require pipelines to generate and communicate explanations. This requirement requires more computation power and resources \cite{ brik2022deep }. ML and AI models use real-time data from the RAN to monitor the RAN's health and performance. As the obtained results enhance O-RAN's security and management capabilities, added costs are justifiable. XAI techniques applied to those ML techniques will require additional time, effort, and resources. Non-real-time \gls{RIC} will require additional computation power to host training jobs for ML models and XAI methods. However, explanations are typically urgent. Therefore, a certain leeway in power is possible. 

\subsubsection{Summary}
RAN commercialization is headed toward an alliance between CRAN, VRAN, and ORAN (xRAN) technologies. Each of these technologies is closely coupled with intelligent systems in operations such as resource allocation and optimization. AI/ML-powered zero-trust architecture will revolutionize security in RAN technologies, from automating user access control policies to auditing. Backing up such integral tasks with a canopy of user-comprehensible explanations would increase the accountability of the intelligent systems used under the hood. 

\subsection{Security of B5G Edge Network} 

\label{sec:edgenetworks}

Edge computing means performing computations near the resource-constrained devices where data is generated as feasible rather than at much further distances \cite{satyanarayanan2017emergence}. Edge layers preprocess data acquired from many sources using caching and processing modules to deliver near-real-time replies to mobile consumers. Because of their advantages in cost-effectiveness in data usage \cite{floyer2015vital}, privacy improvements, and bandwidth usage \cite{murshed2020resource, wang2018bandwidth}, edge networks are becoming increasingly popular \cite{shi2016edge}. Therefore, Edge computing is a widely proposed model for trustworthy AI in the B5G era. \cite{murshed2019machine} Our focus here aims at the impact on the security of edge computing with the advancements of XAI.

\subsubsection{Possible Security Threats, Challenges, Issues}

AI security in B5G edge networks is of two folds, "AI for edge security" and "security for edge AI" \cite{porambage2019sec}. The prior refers to AI techniques used in securing edge systems, while the latter refers to the security of AI systems deployed in edge networks. Also, the authors state that the \gls{DOS} attacks, service or resource manipulation, privacy leakage, and man-in-the-middle attacks are the most prevalent security concerns on edge infrastructure.   

Current research describes artificial intelligence as a facilitator of edge security in various contexts, including general applications and complete architectures that rely on AI. AI4SAFE-IoT \cite{haddadpajouh2020ai4safe} is one such example. The three-layer (network, application, and edge) architecture uses a cross-layer AI engine for security. In that sense, a network layer ML-IDS could mitigate sinkholes, DoS, rank, and local repair attacks in the proposed architecture. The security risks associated with AI in Edge may be reduced by providing the AI modules with more interpretable and fail-tolerant methods that make the models more transparent.

\subsubsection{How XAI can help to mitigate these attacks/issues}

A fundamental security parameter in Edge is the visibility of the whole network to administrators. XAI will be able to achieve this requirement when ML-IDS-based tools come into play. Some designs, such as AI4SAFE-IoT \cite{haddadpajouh2020ai4safe} proposed in the edge layer, may contain various AI models and topologies, which can result in a large number of complex computations performed hidden from the security operators. A proxy model that infers similarly to the actual model but is comprehensible to humans (as in \gls{LIME} \cite{ribeiro2016should}) acts as an intermediary to understand the security parameters' behaviour. For example, explanations can provide information on essential security parameters such as protocol level security, vulnerability towards injections/poisoning/botnets, SLA validity, firewall success rates, and many features that AI/ML models consume. In order to identify backdoor inputs, Hou et al. \cite{hou2021mitigating} offer a filter system based on a mix of classifiers and XAI models. The models can be trained on the server-side edge computers and then sent to each \gls{IIOT} application for identifying backdoor input data, which is then cleaned using an appropriate method. As a result of combining this technique with XAI, the authors claim to have obtained very high rates of backdoor recognition. 
XAI methods reduce the possibility of data pollution/interception, ensuring overall fail-tolerant systems without losing the proprietary system information. Since \gls{LIME} is model-agnostic and comparatively faster, this technique may provide solutions for various applications deployed in edge services such as health, transportation, and agriculture. Research suggestions are also in explainable recommender systems that would be more resilient in place of Edge AI \cite{vultureanu2021recommender} recommender systems as given in \cite{su2019edge}.

\subsubsection{Added cost of Using XAI }

XAI broadens the horizon of edge intelligence (edge caching, training, inference, and offloading \cite{ xu2020edge }) by adding a fifth dimension: edge explanations. Additional costs can be incurred for optimizing the XAI methods for the edge by making them less computationally complex or offloading the computations. Offloading explanation generation with one of the following strategies: device-to-cloud (D2C), device-to-edge (D2E), device-to-device (D2D), hybrid architecture, and caching will address the issue of in-house resource limitations. Access to edge caching for generating and storing pre-model explanations is necessary to ensure security in the edge and IoT layers. 

\subsubsection{Summary}
Edge computing is a powerful tool for reducing costs, latency, and bandwidth usage. However, it also introduces new threats such as MITM, DoS, and privacy leakages. The good news is that AI/ML is increasingly being used to mitigate these attacks, reinforcing systems with a concrete interpretable data flow. This use of AI/ML, combined with the importance of local and global XAI methods, is crucial for improving users' trust in the services, despite the added resource utilization in the long run. 

\subsection{Security of Core and Backhaul Networks}
\label{sec: core-network}

A core network is a highly functional communication facility that links primary nodes and provides communication routes between subnetworks. The backhaul network links BSs to network controllers within a coverage region, which interconnects to the core network through the core transport network. Backhaul network, which is also known as the first mile and last mile (first mile from a fixed perspective, and last mile from a mobile perspective) \cite{chundury2008mobile} is an important part of the wireless data-driven X/AI life cycle. Thus, our focus mainly lies on the intensely AI-used NWDAF and other components that will be playing a major role in the 5G and beyond networks. We discuss the security issues and the impact of XAI in network function analysis components and associated areas.

\subsubsection{Possible Security Threats, Challenges, Issues}

Integrating NWDAF (Network Data Analytics Function) brings a transformative paradigm shift in the core network. The NWDAF is a Network Function (NF) that supports the activities of other 5G and beyond control plane NFs with AI and XAI, making it an integral part of the security landscape. Core network functions (NFs) such as Access and Mobility Management Function (AMF), The Session Management Function (SMF), Authentication Server Function (AUSF), Unified Data Management (UDM), Policy Control Function (PCF), and Network Slice Selection Function (NSSF) are some integral NFs that uses AI models in NWDAF \cite{manias2022nwdaf}. Explanations from each of these components would significantly enhance the accountability and trustworthiness of future networks. Eavesdropping and DoS attacks could take place in the core and backhaul networks. 
Some solutions, such as mutual authentication, key exchange, and perfect forward secrecy, are discussed in \cite{sharma2018secure}. The authors of \cite{liyanage2014case} suggest IPsec tunnel mode and IPsec bound end-2-end tunnel (BEET) mode-based solutions to LTE-backhaul-related security challenges like DoS, virus dissemination, and unwanted VoIP communication. In another paper \cite{liyanage2012secured}, they have addressed TCP reset threats, DoS, and DDoS while proposing a VPN-based architecture for backhaul security. These attacks are far more deadlier in the core networks. However, the solutions with XAI-backed IDS and firewall solutions can immensely help reduce the fallout of attacks.

There is a growing trend of using reinforcement and machine learning methods for backhaul and core network functions. For instance, in \cite{tong2000adaptive}, a Q-learning method is proposed for increasing the dependability of a millimetre-wave (mmW) non-line-of-sight small cell backhaul system. Additionally, the authors of \cite{malila2017intelligent} have addressed the issue of adaptive call admission control using a Q-learning algorithm. The importance of Explainable AI (XAI) for security cannot be overstated in those attacks, as XAI provides critical insights into the decision-making processes of these AI systems, helping to identify, understand, and mitigate potential vulnerabilities and adversarial threats. Authors of \cite{abdulkadir2019optimizing} have emphasized the usage of ML in an \gls{SDN} environment. They have used ANN methods on top of IP routing to estimate and reallocate available network resources to newly added slices using Traffic Engineering (TE) logic. Adversarial attacks on such models can cause disruption in traffic management and the availability of a healthy backhaul network. Despite these works not discussing the interpretability of the used models in real-world applications, the accountability of these tactics and false tolerance is critical. A generic XAI system may be suitable to address this research requirement. The following section explores some of these options. 

\subsubsection{How XAI can help to mitigate these attacks/issues}

Since NWDAF is the main component solely responsible for data analytics and network learning, we will focus on the effect of XAI on its security. In the 5G core, the data required for analytics and logic function of NWDAF are obtained mainly through NFs, including control and security-related NFs (e.g., PCF, NSSF, IDS) and UEs. Thus, any corrupted data could affect the analytics used to monitor the networks. Industrial NWDAF implementations provide closed-loop automation for third-party NFs, which would continuously monitor network slices and UEs under various KPIs \cite{chouman2022towards}. Any continuous data poisoning in the collection phase could severely disrupt the ML-based decision-making process in NWDAF. XAI methods are invaluable in identifying the shift in the AI/ML model's decision-making process over time. In our tutorial\ref{tutoriallink}, we have provided evidence for this shift with results. It can be either done during the data collection phase from the NFs or the validation of the AI/ML algorithms before model updates. The identification process can be automated with explanations in the loop for early prevention of data poisoning attempts. Thus, it prevents false analytics from reaching NFs.
In addition to that, network security functions that involve AI/ML-based IDS would require explanations to guarantee their accountability. However, when moving beyond 5G, the NWDAF must consider new specifications for learning at the edge with low-power requirements \cite{chouman2022towards}.

Industrial implementations can use explainability to improve the credibility of the applications in the face of attacks using explanations for AI-defined actions. The \gls{MLP} used in \cite{abdulkadir2019optimizing} for core network optimization is vulnerable to manipulation with different poisoning methods. However, various post-hoc techniques for explainability, such as case-based reasoning (CBR) \cite{nugent2005case}, coupled into the MLP networks, would provide the required details to delegate responsibilities during malfunctions. In turn, it will reduce reconciliation time and avoid exfiltration attempts. ML-based complex systems are subject in use cases such as policy-based communication in mobile backhaul \cite{torroledo2018hunting, liyanage2015security}. These models are vulnerable to backdoor attacks with poisoned training data. Identifying which samples triggered a backdoor is relatively straightforward from a security operator's perspective. However, identifying the poisoned samples that created the backdoor from the training data is highly tedious, if only possible, with explainability. With XAI, security operators can more comprehensively find out the samples that caused the poisoning with sample importance-based XAI methods. Such security becomes paramount with software-defined monitoring networks in the mobile core and backhaul networks. Automating contingent actions in core and backhaul, where low latency and high throughput are key performance measures in the B5G era, is pertinent. XAI provides the means for doing so in threat analysis and feedback. 

\subsubsection{Added cost of Using XAI }

AI/ML-based systems have been widely adopted in recent studies for dynamic resource management in wireless backhauling. Consequently, XAI techniques become a requirement to increase their resilience and accountability. Deploying XAI methods in energy-efficient small cell backhauling techniques in \gls{UAV}, high altitude platform stations, and satellites \cite{tezergil2021wireless} will be highly challenging regarding costs. These costs might be incurred in addition to computation power, caching, and bandwidth to generate and communicate explanations. Although cost constraints could damper XAI in core and backhaul networks, system insights gained through explanations are important because they indicate that wireless backhaul can be used in the field without any performance losses. They also highlight the adjustments needed for optimal field use and robustness of the AI/ML methods.

\subsubsection{Summary}

Security is a vital component for SDN and NFV-based backhaul traffic monitoring. Better network optimization, architectural enhancements, and security enhancements are envisaged in future research in B5G networks. AI/ML-based systems to identify common attacks such as viruses, MITM, replay attacks, and DoS attacks will also be applied in the core and backhaul parts of the networks. Interpretable AI/ML models are more beneficial than black-box models in avoiding backhaul bottlenecks, balancing the load, and measuring the overall performance of resilience in backhaul networks. We have also summarised the details of this section in Table. \ref{tab:network_layer_table}

\newcolumntype{b}{X}
\newcolumntype{s}{>{\hsize=.15\hsize}X}
\newcolumntype{m}{>{\hsize=.25\hsize}X}
\newcolumntype{n}{>{\hsize=.3\hsize}X}
\newcolumntype{d}{>{\hsize=.18\hsize}X}
\renewcommand\theadalign{t}

\begin{table*}[]

\caption{A general mapping of AI methods and XAI applications for various security tasks in the network layer}
\label{tab:network_layer_table}
\begin{tabularx}{0.98\textwidth}{sndX}
\toprule
\textbf{B5G Enabler} & \textbf{Security Application} & \textbf{\thead{Potential AI\\ techniques}} & \textbf{\thead{XAI support applications}} \\ \midrule


CRAN & Prevention   of multi-stage jamming attacks in CRAN BBU pool  & SVMs MLP, KSVMs (Kernel   trick)  & Aids in creating System threat models \\

ORAN & Malcious xApp detection in 6G & MLP, DL, RF  & Enhancing the detection confidence, closed-loop integration of monitoring applications \\

ORAN & Unseen adversarial attack prevention in 6G xApps & VAE, GANs, XGBoost  & Transparency of detection samples and logical backing, Enhancing the detection confidence, Standardization and analysis during recovery\\

ORAN & adversarial attack creation/Pen testing & Black-box AI models & Model Agnostic Explainers such as LIME and SHAP can be used for reconnaissance or blue team testing\\

Edge Network & Massively distributed DoS attacks & DDoS early detection AI models MLP, XGBoost, AdaBoost & False positive reduction with explanation-trained models, Creation of explanation databases for legal problems\\

Edge Network & Sinkhole DoS, rank, and local repair attacks in 6G massive IIoT & AI4SAFE-IOT, MLP, RF, XGBoost & Improve user confidence with explanations, Accountability analysis, In-model explanation for granularity\\

Edge Network & Zero day adversarial attacks on Edge devices and \gls{CPE} & Small ML models, RF, DT, MLPs & Anomaly detection models reinforced with explanation-based feedback training, post-attack analysis, enhanced monitoring\\

Core and Backhaul & Increasing dependability of mmWave in B5G & Q-learning and other RL methods & Feature importance analysis, Decision path tracking, policy visualization with XAI \\

Core and Backhaul & Adaptive call admission in 5G & Q-learning and other RL methods & Action and reward attribution analysis, counterfactual analysis for admissions, policy comparisons with various users\\

Core and Backhaul & Reallocation of network resources to slices in 6G & MLP, RL & State representation analysis, Action-routine explanations for various resource constraints, policy analysis, In-Model local explanations for false positives and negatives\\

\bottomrule

\end{tabularx}

\begin{minipage}{12cm}
\vspace{0.1cm}
\small MA - Model Agnostic, MS - Model Specific
\end{minipage}

\end{table*}

\section{Role of XAI on Security Issues of Cross Layer aspects}

\subsection{Security of B5G E2E Slicing}

\subsubsection{Possible Security Threats, Challenges, Issues}

Network slicing means partitioning network architectures into virtual elements across a single physical network. It allows operators to meet customized client needs \cite{studios2020}. It is highly analogous to dynamically allocating computer resources to enable concurrent execution of threads in a complex software system, a notion known as program slicing. Program slicing divides (disaggregates) software routines into many threads and configures computing resources to create virtual computing environments for parallel processing. Similarly, through the segmentation of network designs into virtual components, \gls{SDN} and \gls{NFV} provide much more network flexibility than previously possible on top of the physical infrastructure, customizing the deployment of B5G resources and functions required to serve specific consumers and market groups by the network operators.

Authors of \cite{cunha2019network} have reported on both classical (well-researched) security threats and non-trivial (less researched) threats affecting network slicing. Some classical security threats can include traffic injection into interfaces, network slice manager impersonation, host platform impersonation, and unauthorized monitoring of interfaces. Among the non-trivial security threats that are yet to be further researched, passive side channel, active side channel, compromise of the function, and other end device vulnerabilities can be seen as prominent. It is worth noting that these security threats violate at least one of the leading security principles of sub-networks (confidentiality, authentication, authorization, availability, and integrity).

There are numerous instances of AI usage in E2E slicing in various layers (e.g., RL in RAN \cite{raza2019reinforcement}, \cite{bega2019machine}). Here, we scope the management of all these applications in the B5G era. The E2E slicing paradigm entices high-value stakeholders in multiple areas. Also, the ML functions these models address directly affect turnover. For example, In \cite{bega2017optimising}, Q-learning was used to solve the issue of slice admission control for revenue maximization. Also, they have proposed an online Machine Learning-based admission control algorithm that maximizes the infrastructure provider's monetization. These ML models must be secured from attacks such as backdoor attacks. Failure to recognize the attack can directly affect the system's credibility with stakeholders, leading to a massive loss of revenue. The effect of a compromised slice controller can also resonate through slice subnet management functions. If an attacker acquires a token, impersonates a component in the network, or injects traffic to any interfaces,  the integrity and availability can be affected in the controllers or slice managers \cite{cunha2019network}. Detection techniques for these attacks are increasingly drifting towards using ML-based systems. In \cite{tonini2020network}, researchers have tried ANNs, a One-Class Support Vector Machine (OCSVM) based semi-supervised learning model, and a Density-Based Spatial Clustering of Applications (DBSCAN) based unsupervised learning model for in and out-of-band jamming attacks and external polarisation attacks in optical network slicing. Because of these reasons, the accountability and resilience of ML models become critical.

\subsubsection{How XAI can help to mitigate these attacks/issues}

E2E slicing is a paradigm that dynamically optimizes the network by design to thrive, making it an ideal ground for XAI models. Explanations will provide the way ML models perceive essential telemetry data such as protocol details, service feasibility information, SLA requirements (compliance), and host status when providing security-related network functions such as ML-based firewalls, \gls{DPI} and \gls{IDS} by the slice security manager. In return, it will ensure the continuous availability and robust performance of the slice control functions through human user interference or fully closed-loop operations. XAI-based monitoring of the models in \gls{NFV} management and orchestration units \cite{ETSIGRNFVEVE012} can expose features (e.g., vulnerable ports, packet sizes, error rates) recognized by ML-models as more contributing features. Operators can use these for threat isolation even during the early stages of reconnaissance and weaponizing. Either way, constant feedback is required to enable the control loop to enforce policies in mitigating future attacks. For example, in the event of a successful evasion attack on an ML model deployed in the slice manager, the attacker might still exfiltrate, leaving the adversarial samples in the system. These samples are essential to safely create local explanations and generate security policies in the slice security manager to mitigate future intrusions proactively. A quantitative interpretation from the XAI methods obtained in the form of feature/sample importances LIME\cite{ribeiro2016should}, SHAP\cite{lundberg2017unified}, Influence function \cite{koh2017understanding}), counterfactuals \cite{verma2021counterfactual}, and case-based reasoning methods \cite{keane2019case} can be highly useful to the internal stakeholders such as slice/security operators and engineers. More simplified and qualitative explanations from the above outputs suit external stakeholders such as service providers, management-level personnel, and end-users to improve their trust.

\subsubsection{Added cost of Using XAI }

The slices should adapt to traffic changes, detect potential security issues, and take countermeasures autonomously \cite{ tonini2020network }, in each sub-net. For the smooth operation of slice managers, data storage facilities will have to store explanations since real-time generation can be costly. Additional communication protocols must be used to abstract and communicate domain-specific information for explanations alongside interpretations of ML models. Although the GPU optimization required by XAI methods is preliminary and available in a limited number of libraries like SHAP, we can expect it to change very soon with the rapid increase in demand for these methods. Generation of explanations using GPUs from neural networks using large amounts of synthetic data is ongoing research with companies like Nvidia \cite{nvidia2022xaisynthdata} at the moment. GPU-based computing power will be vital for commercial implementations in both ML and XAI. 

\subsubsection{Summary}

While E2E slicing provides the intended network flexibility and adaptability, virtualization adds new vulnerabilities. AI/ML-based security measures can be expected in federated architectures' inter-slice security and network resource harmonization. Previous studies used multiple ML techniques that beg for interpretability at the slice level. This requirement can also be extended toward holistic explanations at the amalgamated domain level. 

\subsection{Security of B5G Network Automation/ZSM}
\subsubsection{Possible Security Threats, Challenges, Issues}

\gls{ZSM} is where the orchestration of cutting-edge technologies like end-to-end network slicing, cross-domain service orchestration, and automation comes together to achieve full network automation. The ultimate automation goal in B5G is to create fully autonomous networks that can self-configure, self-monitor, self-heal, and self-optimize without human involvement. These characteristics need a novel horizontal and vertical end-to-end architecture for data-driven machine learning and AI algorithms. For self-managing AI functions, the ZSM framework depends on \gls{SDN},\gls{NFV} technologies as well \cite{etsi2019zero}. For example, ZSM plans to use DL to provide intelligent network management and operation skills such as traffic categorization, mobility prediction, traffic forecasting, resource allocation, and network security \cite{zhang2019deep}. It introduces a new threat surface that needs to be addressed separately. 

In \cite{benzaid2020zsm}, a range of possible attacks in the threat surface of ZSM on various network aspects is discussed. The E2E service intelligence offered by the ZSM enables decision-making and forecasting capabilities. Consequently, an attacker may design inputs to cause the machine learning models in E2E service intelligence services to make incorrect choices or predictions, possibly resulting in performance degradation and financial loss. On the other hand, this can jeopardize SLA fulfillment and security assurances. Furthermore, API-based attacks such as parameter attacks, identity attacks, MITM, and DDoS attacks; Intent-based interface threats like information exposure, undesirable configuration, and abnormal behavior; threats on closed-loop automation control systems such as deception attacks; AI/ML system target attacks such as poisoning attacks and evasion attacks; threats on Programmable Network Technologies such as DoS, privilege escalation, malformed control message injection, eavesdropping, flooding and introspection attacks are some of the attack vectors emphasized in the threat surface of ZSM.
On the other hand, major entities such as governing bodies, investors, and researchers must make deliberate decisions on policy standardizations and applications to ensure intelligent urban development. With closed-loop systems that use black-box AI methods in a significant proportion of operations, explanations become necessary to maintain the transparency of the decision processes and enable the governing bodies to reach fruitful policies and regulations. The use of XAI techniques in intelligent monitoring systems, which autonomously collect, analyze, and communicate data to maintain automation, is important in these scenarios. \cite{luckey2020artificial} notes the usefulness of techniques like LIME (Local Interpretable Model-agnostic Explanations) and LRP (Layer-wise Relevance Propagation) for providing transparent and understandable insights, which are essential for maintaining the security, reliability, and trustworthiness of automated network operations.  

In literature, \cite{benzaid2020ai}, the authors emphasize challenges such as the need for AI/ML security and how AI model interpretation will guarantee accountability, reliability, and transparency by improving the trustworthiness of AI-enabled systems. However, they also mention that the research gap in the field of ML security for network and service management is limited to only a few contributions (i.e., \cite{usama2018adversarial, han2018reinforcement}). 

\subsubsection{How XAI can help to mitigate these attacks/issues}

Closed-loop network automation requires explanations due to the widespread use of AI/ML-based network functions throughout the architecture. ZSM relies on trust-based relationships among the diverse management functions. These context-based AI/ML trust models will generate triggers based on their security assessments \cite{ETSIGRZSM010}. Explanations will be essential to verify the security assessments required to quarantine the management functions. For example, the XAI would give insights about the operational status/changes, network status, package versioning, consumer information, and subnetworks compromises as seen from the AI/ML model’s angle \cite{ETSIGRZSM010}. Such explanations possible to be derived through Partial Dependence Plot (PDP), Individual Conditional Expectation (ICE), Accumulation Local Effects (ALE) Plot, Feature Interaction, Feature Importance, Global Surrogate, Local Surrogate (LIME), and Shapley Values (SHAP) \cite{lundberg2017unified, islam2021explainable}. In addition, explanations can provide context on how AI/ML models use stakeholder information (handling multi-tenancy) in management domains. For example, the tenant ID, tenant-specific security/isolation/access policies, will be used in the decision-making process of AI/ML models, making them opaque during internal processing. However, with XAI methods, the reasoning in these models will be transparent to the outside \cite{ETSIGRZSM010}.

IBN is a network that operates autonomously with the intent of a predetermined set of directives. Unlike an imperative policy, an intent-based policy is a set of objectives that must be completed throughout network operation to achieve collective performance goals. XAI is the cornerstone in realizing intent-level trust, given the contextual awareness and appropriate data from multiple networks and intent functional blocks. Service orchestration optimization, resource monitoring, context and behavior-based intent to service mapping, and extracting service primitives from intents are some of the operations where explanations might be important \cite{ mehmood2021intent, ETSIGRZSM005}. Malicious agents will manipulate such parameters in the process of attacks. However, the operators can easily monitor them using explanations to handle incidence response, contingency planning, and risk mitigation.

\subsubsection{Added cost of Using XAI }

Several management tasks are bundled together in the ZSM \gls{MD}, such as domain data collection services, domain analytics services, domain intelligence services, domain orchestration services, and domain control services \cite{ benzaid2020ai}. Additional channel bandwidth to communicate explanations generated about domain intelligent services and data collection services must be looped into the domain analytics of each MD so that any changes required in domain control and orchestration are adequately executed. GPU and CPU computational power for XAI will also be an added cost. Furthermore, generated domain-specific explanations must be stored in each domain data service. In contrast, cross-domain explanations will be stored in standard data services, calling for additional storage and caching space. The existing architecture can be conveniently adapted to explanation-based analytics with minimal compromises. 

\subsubsection{Summary}

ZSM or network automation can be simply identified as the future of telecommunication systems. In full automation, AI/ML is integral to the closed-loop management of a network. In closed-loop management, an undesirable configuration or an attack on AI/ML-based systems can pull malicious behavior into a whirl of abnormalities in the network domains. XAI is a viable candidate for uncovering any underlying AI/ML systems vulnerabilities and shedding light on obscured attack data in black-box models. 

\section{Role of XAI on other security enablers in B5G} \label{sec:xai_role_other}

In this section, we examine the role of XAI and its adaptability in time-tested security techniques. The discussion includes the interoperability between XAI, encryption, and distributed systems brought into the 5G and beyond era.

\textit{XAI applied in the traditional security methods} would significantly affect the perception of services provided in the B5G era. Novel technologies should complement the existing technologies and flawlessly fit into the existing systems. On the other hand, XAI shows a giant leap ahead of traditional security methods. XAI addresses the security objective of accountability (for AI/ML), unlike the traditional CIA triads (Confidentiality, Integrity, and Availability defined in NIST (National Institute of Standards and Technology) standard FIPS (Federal Information Processing Standards Publication) 199). In this regard, XAI helps to deobfuscate the black-box nature of ML methods (used for security or not) and trace the problem(security-related or malfunctions) to responsible parties/attributes. Therefore, XAI methods fill the research gap of narrowing down to the exact constituents of AI systems that malfunctioned/compromised during security audits rather than broad replacements. 

In future networks, standard security approaches such as access lists, encryption methods, and distributed systems alone cannot provide the benefit of detecting specific information about the AI models. Nevertheless, conventional methods may enhance the security of explanations produced in the context of 5G and beyond networks.
In the following subsections, we discuss how XAI would use the existing security mechanisms and enable smooth deployment alongside them. 

\subsection{XAI and encryption methods}
Encryption will be continued to B5G \cite{sakamoto2021rocca}, and it is one of the best privacy-preserving security methods in networking. It includes diverse techniques such as data obfuscation, cryptography, and data anonymization. Encryption is a heavily discussed topic regarding privacy-preserving AI and, therefore, weighs considerable importance in applying XAI to said AI methods. Applying XAI to encrypted data will be a challenging task that has yet to be explored. In the XAI pipeline, there are three points where encryption could create an added layer of security. (1) Input data where X/AI is trained on (2) The XAI model, which uses encrypted input data/encrypted model, and (3) explanation encryption. However, encryption will also require additional computation power.

\subsubsection{Data obfuscation}

Data obfuscation is expected to be used with B5G services to maintain the privacy of systems \cite{pan2021differential, wan2022privacy}. It refers to changing the data communicated in the network to confuse the counterparties trying to intercept the data. It can be done on either input data for the XAI method. The current literature includes additive \cite{liu2005random} and multiplicative \cite{xu2019lightweight} perturbation-based obfuscation methods. Popular techniques such as \gls{DP} fall under additive perturbation-based obfuscation. The use of DP in wireless data transfer mainly occurs on the perception layers extending up to the service layer. In use cases such as Smart Health and Industry 5.0 in the B5G era, data in all stages of the life cycle have to be secured with one or more forms of obfuscation due to their critical and sensitive nature. Nevertheless, one can argue that it could be more helpful at the decision stage than the output stage since a particular explanation should be quickly understandable to human users (Similar to \cite{gilad2016cryptonets}). In a model trained with obfuscated data, the explanations must compensate for the obfuscation and reflect only the contribution of the data to the outcome. The perturbation-based XAI methods can further obfuscate perturbation from obfuscated data. This incident could lead to false explanations and false security alerts. Another option would be to carry out deobfuscation before applying pre-model XAI. It can expose the pipeline to leaks if not done correctly. Current research is minimal in making a concrete claim on this scenario.

\subsubsection{Cryptographic methods}

Although cryptography will undergo significant changes (and challenges with quantum computing) in the B5G, it is still expected to prevail for the long haul \cite{bhatt2023post}. It is the science of creating cipher texts from plain text, encrypting the data, and making it completely unreadable directly to humans. Generally, Cryptographic methods are heavily involved in ensuring security in networking protocols. 
Furthermore, cryptographic methods have extended towards enabling computation on algorithms utterly oblivious to the data. It is essential as the third-party cloud servers running X/AI/ML models can not be fully trusted unless they are in a \gls{TEE}. \gls{HE} and \gls{MPC} are popular encryption methods proposed in the literature to ensure the privacy of ML models in the cloud. However, they are required to do calculations on encrypted data in an oblivious way. Work with XAI on these methods is minimal currently. Apart from data oblivious hardware implementations such as oblivious RAM \cite{mitchell2014data}, which in general could be helpful, there haven't been any studies that are currently useful to apply oblivious computing for XAI specifically. \gls{HE} encrypts the data and generates a key pair, one private and the other public. The public key and instructions will be available to third parties. XAI models can use this key and instructions to generate explanations. However, there is a gap in the literature for a rigorous analysis of applying popular XAI algorithms (LIME, SHAP) on \gls{HE} based models.

\subsubsection{Data anonymization}

Data anonymization will be paramount in the B5G era, considering the personal data collected through various personal equipment such as wearables \cite{silva2022privacy}. Data collectors must remove user-sensitive data fields such as name, sex, and ID from the dataset, which would otherwise expose personal information about the users. For models in the context of B5G, excluding extraneous features like personal names and demographic-related characteristics could prove mutually beneficial for both ML and XAI methods. An anonymization technique such as k-anonymity \cite{sweeney2002k} would remove the explicit identifiers and balance the distribution based on other identifiers. This method is not entirely foolproof. The effect of removing these features is quantifiable with feature importance-based XAI methods. These metrics can help operators recognize the actual cost of anonymization and add or remove unimportant features for the inference process. It ensures higher data privacy in the face of model inversion attacks.

\subsection{XAI and Federated Learning}

Distributed edge computing with federated learning is a prominent field of study in the B5G era, as Wan et al. (2022) highlighted in their work on privacy concerns \cite{wan2022privacy}. The use of Explainable Artificial Intelligence (XAI) within the framework of federated learning approaches has potential benefits in the realm of security for Beyond 5G (B5G) networks, owing to many factors. FL's critical applications in areas such as automated vehicle networking \cite{renda2022federated} and edge communication \cite{lim2020federated} (Section \ref{sec:edgenetworks}) are one of the main reasons. However, FL's threat landscape could expand due to the high connectivity of heterogeneous devices. Although FL provides higher privacy, detecting malicious agents among them can be tedious without a transparent technique. Thus, explainable systems will ensure the accountability of FL in the future.

Aggregation modules are predominantly a cloud/edge-based approach, which relies on the deployed environment. If the FL algorithm is deployed inside a \gls{TEE} the aggregator, the aggregator can be assumed to be secure. The threat vectors can originate from the clients and respective communication channels in different attacks, such as poisoning. However, if the execution environment is untrusted, FL can be exposed to attacks in the aggregator and the clients. That would be rather severe in damage. Premonition and quick isolation of malicious actors in FL is possible with attribution/sample-level explanations. For example, poisoned data from a health monitoring device can be identified using feature-based XAI methods. Explorative data analysis could also help to recognize out-of-distribution data points. It can be helpful to trace down the malicious clients in the network and take suitable actions. If the aggregator is in an untrusted environment, monitoring systems with XAI methods should be used to track any changes in the aggregator. SHAP is a promising technique for observing the overall shift of the model through feature importance values. It is a viable detection mechanism against evasion attacks in ML models \cite{fidel2020explainability}. In the B5G era, this attack is highly probable and can be used to detrimental effect.

\subsection{XAI for LLMs in future networks}


 \begin{figure}[htb]
    \centering
    \includegraphics[width=0.40\textwidth]{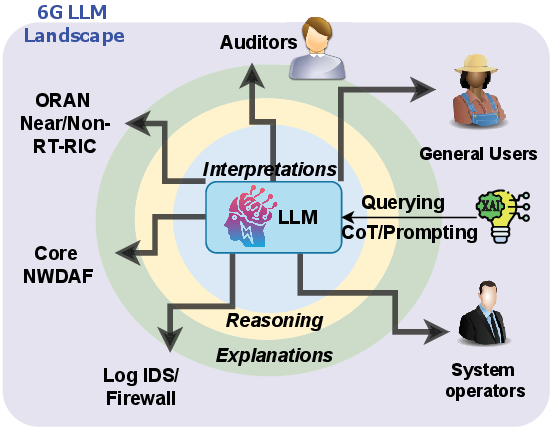}
    \vspace{1mm}
    \caption{LLMs would require more and more transparency in the coming days to use their full potential in critical applications. However, even the current application potential shows that LLM and XAI integration would mutually benefit the fields. Explanations would be improved by LLMs and in return LLMs can also be made fairer for both human and autonomous applications.}
    \label{fig:llms}
\end{figure}

\gls{LLMs} have been gaining attention in the past few years and practically have exploded in popularity (gained popularity in 2023, although ChatGPT was introduced in 2022) with ChatGPT-3 from OpenAI. Since then, various LLM-based applications have been introduced. Google's PALM, Gemini and Bard, and Meta's LLaMA are popular examples with billions of parameters \cite{ long20246g }. LLMs in 5G and beyond are anticipated to strengthen security measures of the networks via LLMSecOps \cite{ nguyen2024large }, which prioritize safeguarding and optimizing network operations along the 6G edge-cloud continuum. We would like to direct the readers to literature such as \cite{ xu2024large, long20246g, zhou2024large, lin2023pushing } for further reading on LLM in 5G and beyond networks. Using LLMs, 5G and beyond networks may efficiently include security features such as IBN, NWDAF, and zero-touch network security. This integration seeks to establish higher standards for safe and intelligent network management, creating a foundation for a stronger and more resilient infrastructure. Furthermore, LLM-based penetrating testing tools (e.g.: PentestGPT \cite{ deng2023pentestgpt }), anomaly detection tools ( e.g., PAC-GPT \cite{ kholgh2023pac }, LogBERT \cite{ guo2021logbert }, LogBot \cite{ balasubramanian2023transformer } ), and threat detection tools (e.g., TSTEM \cite{ balasubramanian2024tstem }, Cyber Sentinel \cite{ kaheh2023cyber })

The rise of this new addition to the security stack of 5G and beyond begs the question, "How secure are these components in deployment?". Not only in the security of the above models but also in various other aspects of 6G (as shown in Fig. \ref{fig:llms}) where LLMs are foreseen to be used such as network optimization, intelligent control operations in the 6G integrated TN-NTN with IoT services \cite{ rong2024leveraging }, and LLM enhanced \gls{RIS} \cite{ liu2024llm } requires accountability and trustworthiness. LLMs are larger in number of parameters and notoriously black-boxed, so the interpretations can be highly challenging on their own. If not interpreted properly, they can lead to harmful content, such as in Google AI overview \cite{Grant_2024} or hallucination \cite{ weidinger2021ethical }. In this regard, XAI is a formidable tool for enhancing the security of an LLM. 

Explaining LLMs in security can be approached in the general methods of local and global explanations \cite{ zhao2024explainability }, where local explanations would take the form of feature attributions, attention, or examples, as discussed in the previous sections. For instance, in an LLM that uses logs to detect threats in \gls{RIC}, the non-real-time component can isolate the features that have contributed to a false classification or anomaly. Furthermore, in future training rounds, the model can be optimized not to weigh in on those features. With example-based explanations, xApps (Non-RT) can filter out malicious contributions in the training rounds. Popular concepts such as Shapley values \cite{ chen2023algorithms } can also be used similarly. On the other hand, global explanations would delve deeper into understanding the workings of neurons and hidden layers of LLMs deployed in security components. Probing-based explanations \cite{ chen2021probing }, neuron activations \cite{ bills2023language }, and concept-based explanations\cite{ kim2018interpretability } are even more critical in developing IBNs. LLMs used in intent translations require high accuracy to ensure the most secure configurations are created in a closed-loop automation system. Thus, both global and local explanations must be verified for accountability and trustworthiness. However, for explaining larger LLMs and in critical use cases, global and local explanations can be too computationally expensive and unsuitable for various use cases \cite{ zhao2024explainability }. For example, IBN in the 5G and beyond networks requires developments in AI, such as LLMs, to effectively convert such intents into configurations. It will utilize LLMs to enable run-time network configurations through high-level intents to simplify human-network interactions and smoothen the deployment of new services. To ensure the security of the whole process, it is advisable to opt for prompt-based explanations that would consider the reasoning abilities of LLMs \cite{ wei2023larger }. For example, \gls{CoT} explanations would prompt an intent translator in a zero-touch network to translate an intent, resulting in the output of steps: creating an IoT network, assigning IP pools, and assigning firewall rules. Then, the explanations can be acquired by directing these prompts in a specific way and having them explain the reasoning. As the implementations of these design steps need to be as predictable as possible with security, it is imperative that the reasoning is concrete. These techniques can be similarly applied in LLMSecOps services \cite{ nguyen2024large } paired with other techniques such as counterfactual prompting.

\begin{table}[!ht]

\caption{Summary of benefits from XAI in B5G technical aspects}
\label{tab:xaibenefits}
\begin{tabularx}{0.45\textwidth}{lX}
	\midrule
	\textbf{Layer} & \textbf{Benefits of XAI for B5G security} \\
	\midrule
    Perception layer & \begin{itemize}[noitemsep,nolistsep] \item{}Recognize samples manipulated in sensor data poisoning
    \item{}IoT/device explanations help to communicate security status to the user quickly
    \item{}Localizing malicious attributes in malware detection
    \item{}Helps in monitoring device parameters during attacks
    \item{}Obtain contextual information IoT devices to improve awareness
    \item{}Realize packet information misinterpreted by ML models
    \item{}Identify malicious actors' information, such as IPs and other connection telemetry manipulated in devices by intruders
    \end{itemize} \\
	\midrule
    Network Layer & \begin{itemize}[noitemsep,nolistsep] \item{}Enables granular monitoring of inference from xApps, and rApps for malicious actions in ORAN
      \item{}Enables sample-level analysis of ML-based ORAN SMO control decisions 
      \item{} Explanations are useful in generating synthetic data for red team testing of RT-RICs.
      \item{}Increases confidence of recommendation generated in the edge layer when accompanied by explanations
      \item{}Supports creating security policies and conflict mitigation methods based on ML models.
      \item{}Strengthens the traceability of the attacks to respective slices and tenants in core networks.
      \item{}Enables close monitoring of the ML models' features; throughput, latency, and QoS parameters.
      \end{itemize} \\
	\midrule
    Cross-layer aspects & \begin{itemize}[noitemsep,nolistsep]     \item{}More user-friendly human-machine interfaces can be developed for network automation
      \item{}Explanations can be a vital component in the feedback of close-loop domain managers in automation.
      \item{}XAI helps to reveal vital evidence when reverse engineering adversarial samples in policy control for ZSM
      \item{} Interpretations are useful in uncovering attributions made by ML models on tenant information. 
      \item{}Explanations reveal parameters for improving the security of orchestrators and intent-based services perceived through ML models.
      \item{}Explanations will help optimize security policies to slice security managers to make them more efficient.
      \item{}Enables early identification of security backdoors in AI/ML full automation loops.
      \end{itemize} \\
	\bottomrule
\end{tabularx}

\end{table}

\subsection{Summary}
We recognized the following advantages of using XAI to complement traditional security methods. We have listed these according to the defense stages defined in the NIST framework for cybersecurity \cite{NIST2018framework}.
\begin{itemize}
    \item Identifying: XAI models are mostly transparent or interpretable easily. Thus, vulnerabilities, biases, and malfunctions can be easily identified. Unlike using encryption methods for everything, XAI answers the question \textit{Why} do we have the need (vulnerability) to apply encryption methods?
    \item Protect: Alongside access control methods such as access lists, XAI applications can recognize users through their context. Access lists need further reinforcement with the inferences of XAI. However, new XAI protocols must be developed for security purposes, as current protocols cannot accommodate explanations.
    \item Detect Transparent models such as DTs and RFs are used in IDS systems due to their explainability that balances the transparency of rule-based systems and the accuracy of complex DL models.
    \item Respond: XAI is used to recover information on the compromised parts of an affected complex AI/ML. These techniques can answer the question of \textit{What internal function in the AI model affected the security breach?} Encryption methods and access lists are insufficient to dig into AI models.
    \item Recover: Explanations about AI models on the effects are useful in creating recovery plans and restoring trust with stakeholders. Traditional techniques are lagging in this aspect.
\end{itemize}

\section{New Security Issues and Challenges of B5G Due to XAI} \label{sec:new_issues}

Although using explanations for AI and ML solutions has numerous advantages for B5G security, it can also add specific challenges and issues to systems. In this section, we discuss current and foreseeable issues and challenges of XAI and how it would translate into 5G and beyond the era of telecommunication with the ubiquitous use of AI/ML. The most suitable analogy would be a double-edged knife as the attackers also can use XAI to understand how the black box model works, complicate the design process for system architects/developers (i.e., explainability must be addressed in the trade-off between model performance and security), and create new attack pathways (i.e., the explanation itself can be falsified).

\subsection{Increased vulnerability to adversarial ML attacks}
\subsubsection{Introduction}

Many existing attacks target ML models: adversarial ML attacks~\cite{biggio2018wild}. \textit{Membership inference} and \textit{model extraction}~\cite{juuti2019prada} attacks compromise the confidentiality of the training data and the ML model respectively. \textit{Model poisoning} and \textit{model evasion} attacks (a.k.a. adversarial examples) compromise the integrity of the ML model and its predictions. A common characteristic of adversarial ML attacks is their effectiveness increases as the attacker's knowledge about the ML model and its decision process increases. Consequently, the obfuscation of ML models' decision process, by making it a \textit{black-box}, is an effective defense to mitigate adversarial ML attacks~\cite{papernot2017practical}.

\begin{figure}[htb]
    \centering
    \includegraphics[width=0.48\textwidth]{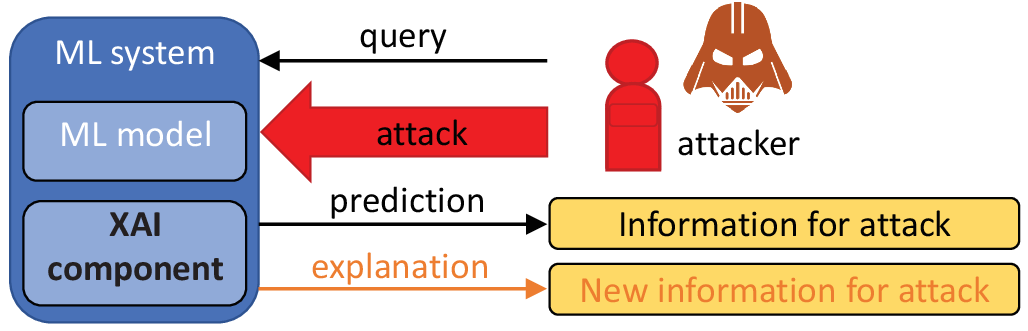}
    \caption{Explainability reveals new information that ``white-box'' black-box ML models and facilitates adversarial ML attacks against them.}
    \label{fig:whitebox}
\end{figure}

In the context of ORAN, core, and edge networks of 5G and beyond, security concerns arise when attackers attempt to deobfuscate the models with explanations. This process can potentially reveal sensitive information about the decision-making process of black-box ML models, as illustrated in Figure~\ref{fig:whitebox}. Explainable ML techniques can inadvertently aid attackers in designing more effective black-box attacks~\cite{kuppa2021adversarial}.

For instance, attackers can leverage the information produced by explanations to enhance their capabilities in membership inference, model extraction, poisoning, and evasion attacks against black-box ML models within these network components. Explanations can provide insights into the crucial features influencing model predictions, allowing attackers to refine their strategies for crafting perturbations and evading or poisoning the models.

Moreover, the explanations accompanying ML model predictions can support attackers in intentionally misclassifying a sample. By identifying essential features that influence the model's output, attackers can iteratively modify these features based on feedback from the model predictions supported by explanations. This iterative process enables attackers to gradually refine their techniques, ultimately evading the models deployed within the networks of 5G and beyond.

\subsubsection{Impact on B5G}

There is currently no 100\% infallible protection against any of these adversarial ML attacks, and the problem of how to defend against them remains an open research question. Some vulnerabilities exploited by adversarial ML attacks can also be necessary features of ML models~\cite{andrew2019adversarial}. Explainability contradicts the usage of obfuscation to make explanations more resilient on the grounds of one of the main goals, i.e., to make the explanations clear and concise to the end-user. This increased exposure decreases the security of some non-critical use cases in B5G.

For instance, ML-based security measures deployed in the 6G sub-networks perimeter monitor the anomalous behavior within the sub-network or reaching from other sub-networks~\cite{siriwardhana2021ai}. DT, Random Forest, DNN, clustering, ensemble methods, and Gradient Boosting Machines (GBM) are used to detect common network attacks, like DDoS attacks, from traffic data~\cite{singh2018ml, amaizu2021composite, siriwardhana2021ai, idhammad2018semi }. The evasion and poisoning of these ML-based anomaly detectors open to adversarial attacks~\cite{ kuppa2021adversarial }. With explanations, this is made more accessible to some degree. One outcome is malicious traffic bypassing the system defenses and exhausting network resources. It can reduce the system resources available to favor legitimate users.

In closed-loop E2E service management, intelligent services use ML-based decisions. When these ML models have transparency or interpretability, it becomes easier to create counterexamples against them. Even though E2E slicing interpretations can have improved encryption/access lists, hostile instances can still lead to incorrect predictions and decisions. It can cause an overestimation of resources needed by a slice in the future or improperly reset the management policies. \cite{benzaid2020ai}. 

\subsubsection{Possible solutions}

A solution to this issue lies in controlling the explainability of provided information. First, one must define the minimum requirement and granularity of the explanation required to achieve an intended goal. The selected XAI method should only meet this minimum requirement without revealing more information than necessary. Stakeholders such as creators of ML models would require the highest possible transparency. Also, the developers would require high interpretability depending on the tasks. However, access to end-point developers and ML engineers must be different, with the latter requiring higher interpretability.  

Second, one must control the access to the explanation, i.e., restrict it to only the necessary parties. The explanation can also be sealed, encrypted, and only revealed if there is a need to investigate a decision of the model, e.g., stakeholders such as auditors, and regulatory bodies. The default access to explanation must be as restricted as possible rather than wide open. This restriction limits the opportunity for an attacker to access this information. 

Finally, delaying the availability of explanation (by a few hours or days) when compared to the availability of the ML model decision can slow down attacks. In many ML use cases, the decision from the ML model must be obtained quickly, while the explanation is not time-sensitive. Adversarial ML attacks are typically iterative, counting 100s of steps. Each new step relies on the information from the previous step(s). By delaying the availability of explanation, the utility of the ML model is not impacted, while an adversarial attack can be drastically slowed down or even completely prevented. Stakeholders such as system operators monitoring real-time systems (e.g., in ORAN RT-RIC) are an exception in this regard. 

\subsubsection{Summary}

Currently, there is no perfect solution to fix the use of XAI methods against ML models. System-level attack prevention is the most effective method against using XAI methods to improve adversarial attacks. These methods include access restriction, the encryption of explanations, or delay in response. This may change in the future as defenses against adversarial ML attacks become effective, and a foolproof defense against some of these attacks would be developed. 

The issue raised here is that current adversarial ML attacks are more effective in a white-box than in a black-box setting. Explainability has the ``white-boxing" side-effect on black-box models.
There is work already showing that, e.g., membership inference attacks can be run as effectively against black-box and white-box models~\cite{sablayrolles2019white}. In such cases, a black-box model explained using XAI (white-boxed) would not be more vulnerable than its non-explained counterpart. The impact of XAI on its security would thus be canceled. 

\subsection{New attack vector and target}
\subsubsection{Introduction}

Post-hoc methods for XAI are new components added to ML-based systems. This new component can complement the prediction of ML models, weighing heavily on the actions of systems and humans that depend on the ML model. In some cases, the explanation itself is more important than the prediction. This is the case for AI used in applications having a societal impact, where predictions must be fair and unbiased. This is also the case for security applications like detection and response (D\&R), where an explanation is used to counter and recover from detected attacks using appropriate measures.

\begin{figure}[htb]
    \centering
    \includegraphics[width=0.45\textwidth]{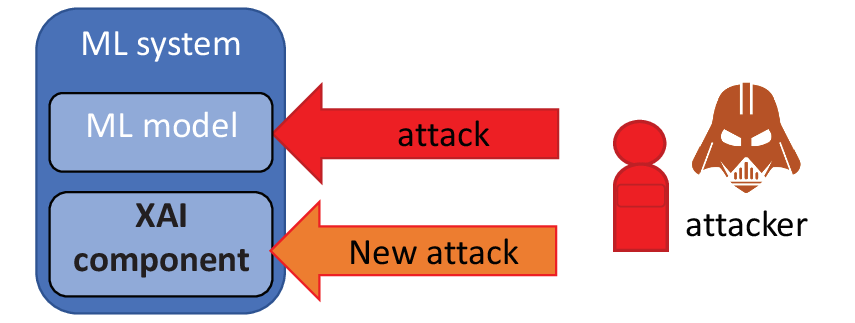}
    \caption{The XAI component becomes a new target and a new attack vector to compromise the whole ML-based system.}
    \label{fig:new_attack_vector}
\end{figure}

Due to the importance of explanation, the XAI component can become the main target of an attack, as depicted in Figure~\ref{fig:new_attack_vector}. Penetrating the RIC of ORAN can be a possibility for a malicious third-party app willing to impact the QoS of the system \cite{porambage2023xcaret}. Such a manipulation can lead to network congestion, latency issues, and compromised user experiences. 
Directly attacking post-hoc XAI methods can change the explanation while the prediction of the ML model remains the same, as demonstrated in~\cite{kuppa2020black, galli2021reliability}. The ML model makes the right decision, but the dependent system or human takes a wrong course of action based on the incorrect explanation.

XAI may also mask biased ML model results with false reasons. \textit{Fair-washing} \cite{aivodji2019fairwashing} is the misperception that an ML model meets particular requirements while its behavior drastically deviates from its justifications. It is demonstrated further that post-hoc explanatory approaches depending on input perturbations, like LIME and SHAP, are unreliable and do not give definitive information regarding fairness~\cite{slack2020fooling}. An interpreter-only attack technique known as \textit{scaffolding} is built based on this observation. An attacker can generate desired explanations for a given unfair ML model (which uses LIME/SHAP) by masking any biases in the model. Through this hack, a compromised XAI method enables hiding biased/unfair outcomes indicating that they are harmless/unbiased. 

\subsubsection{Impact on B5G}

This threat exists for every ML application to B5G, where the explanation weighs equally or more than the prediction in the action it triggers. For instance, in D\&R, where an explanation is used to counter and recover from an attack, modifying the explanation for a prediction leads to creating ineffective safety policies in E2E slice security managers and conflict mitigation in ORAN. 

Certain decisions require sensitive user data to ensure the security and safety of the services provided. It is also essential to explain the data usage in decision-making to the human user. Critical applications such as autonomous driving are envisaged to rely on B5G networks~\cite{chen2020deep}. When a system fails or crashes, the explanation for the incorrect prediction that led to it will be paramount as evidence for the following legalities. The stakeholders of these situations will have to decide whether a system or human error caused the losses. If the explanation gets manipulated, the responsible parties could go without consequences. 

\subsubsection{Possible solutions}

The main reason for this new attack target is that the explanation of post-hoc methods can sometimes be disconnected from the prediction of the ML model they interpret. Leveraging explainability through transparency would provide the explanation to come directly from the ML model itself, and it is usually well linked to its actual decision process. Both the ML model and explanation process must be fooled to succeed in an attack. Even though this is possible~\cite{kuppa2020black}, it is more complicated. Furthermore, by using XAI methods based on transparency, an explanation would be partly protected by existing defenses against adversarial ML attacks that already protect the ML model. The state of security in the prevention of adversarial ML attacks is more advanced than it is for the protection of attacks against XAI methods. 

However, if explainability through transparency is not possible, selecting different post-hoc explanation methods can increase resilience against attacks. For instance, empirical experiments~\cite{slack2020fooling} show that SHAP is more resilient than LIME when it comes to hiding biased and unfair outcomes.

\subsubsection{Summary}
Adding new functions and components in large systems increases system complexity and vulnerabilities, exposing new attack vectors. XAI, primarily through post-hoc explainability, is a new component that exposes new attack vectors against ML-based systems. If an ML model is transparent by itself, for an attack to be successful, it must fool the ML model and its explanations, making it more challenging for the attacker.

\subsection{Challenge to design secure ML applications with XAI}

\subsubsection{Introduction}

 \begin{figure}[htb]
    \centering
    \includegraphics[width=0.32\textwidth]{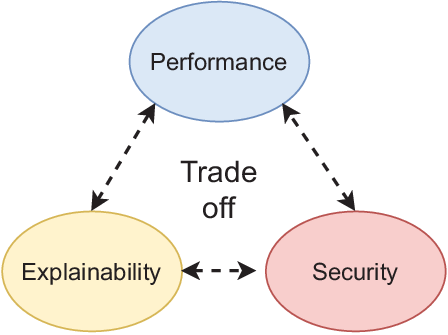}
    \vspace{1mm}
    \caption{New trade-off required between performance, security, and explainability of ML systems.}
    \label{fig:perf_sec_exp}
\end{figure}

The design and implementation of ML-based systems are guided by the sole requirement of maximizing performance, i.e., high accuracy, high generalizability, and low response time. Adding security requirements to ML-based systems introduced the first trade-off between antagonist properties: performance vs. security. It has been shown that effective defenses against adversarial examples, like adversarial training, degrade the accuracy~\cite{kurakin2016adversarial} and the  generalizability~\cite{raghunathan2019adversarial} of protected ML models. There also exist trade-offs between security properties. For instance, increasing the resilience of ML models against evasion attacks makes them more vulnerable to privacy attacks like membership inference~\cite{song2019membership}.

Explainability is a new requirement that is adding to the existing trade-off. Three properties constraining each other need now to be fulfilled by ML systems, as illustrated in Figure~\ref{fig:perf_sec_exp}: performance-security-explainability. A transparent model can have higher explainability and security as a result. Nevertheless, explaining with transparency will limit the number of models one can use. This dilemma potentially leads to discarding the solution providing the best accuracy to meet the explainability requirement.

In the 5G and beyond networks, these trade-offs can also bring forth some other challenges. ORAN systems can suffer from software flaws and insecure designs. These can extend to improper storage of sensitive data and logs \cite{liyanage2023open}. When the security is traded for explainability, those problems can accentuate vulnerabilities. The same principle can apply to core and edge networks as well.

\subsubsection{Impact on B5G}

The new requirement of achieving a performance-security-explainability trade-off makes it challenging to design well-balanced ML systems for B5G edge networks. Deploying ML models on-device enables training using federated learning and local decision-making, making communication more efficient. On the other hand, device resource limitations make running ML models on a device challenging. Performance becomes thus a primary requirement constrained by device resources, relegating security and explainability to secondary places. 

For example, body-sensors/fit-bits collecting vital signals to provide dietary and physical recommendations struggle to squeeze out the necessary computational power to run sophisticated cryptographic techniques on top of ML models, and they fail to provide sufficient security~\cite{siriwardhana2021ai}. These constraints require developers to use transparent low-power models to preserve explainability, which might not be an ideal model selection for the particular use case in terms of accuracy, robustness, or privacy. 

\subsubsection{Possible solutions}
One can use post-hoc explanations instead of in-model explanations as a solution. Nevertheless, this solution has two drawbacks. First, the explanation from post-hoc methods sometimes has a lower correlation to the actual decision of the model, so it offers a lower-quality explanation. Second, the post-hoc solutions create new attack vectors and targets against the whole system, including the ML component. Thus, the vulnerability introduced by post-hoc explainability is moved from the ML-model in the end-device to more trusted environments in the cloud. 

A second solution is the careful analysis and prioritization of the ML system requirements. Evaluating and quantifying the performance-security-explainability trade-off leads to making an informed choice about which requirement(s) to meet and which other(s) to neglect. Requirements neglected during the ML model design may be addressed later at the system level. The security of ML models can be increased through system security, e.g., by detecting adversarial queries to the model at inference time~\cite{juuti2019prada,szyller2021dawn}.

\subsubsection{Summary}
 The \textit{trustworthy AI} concept aims to ease these worries by enforcing a large number of desired properties to make AI and ML applications trustworthy~\cite{eu-ai-act}. Among the first requirements were accuracy, performance, security, and privacy. Many more requirements were added, such as explainability, transparency, accountability, fairness, etc. Thus, the vulnerability introduced by post-hoc explainability is moved from the ML model in the end device to more trusted environments in the cloud. 

\subsection{Ethical and other considerations of XAI usage}
\subsubsection{Introduction}

 \begin{figure}[htb]
    \centering
    \includegraphics[width=0.39\textwidth]{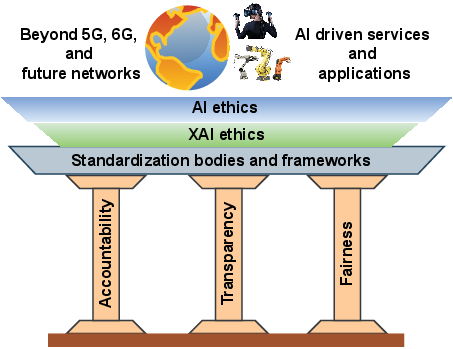}
    \vspace{1mm}
    \caption{The main pillars where XAI ethics and AI ethics rest are shown in the figure. Timely expansion of the ethical frameworks for XAI and AI is expected in the road to 6G}
    \label{fig:ethics}
\end{figure}

The ethics of using explainable AI is an important topic, considering the lengths it would reach in the realms of 5G and future networks. The ethics of XAI we discuss in this section are limited to the application of AI/ML-based security approaches in future networks. However, readers should distinguish AI ethics from XAI ethics, which we discuss here. The concepts of AI ethics and XAI ethics overlap highly, and AI ethics can be considered the umbrella category that includes XAI ethics. AI ethics refers to designing, developing, and deploying AI systems that respect human rights and values, which include principles, legal requirements, and rules that developers and creators should adhere to. As shown in figure  \ref{fig:ethics}) fairness, accountability, and transparency are the central pillars of ethical AI \cite{ vainio2023role }. Transparency pertains to the comprehensibility and clarity of a system, while justice and fairness indicate the principles of equality and impartiality; accountability refers to the individuals held liable for the system's activities. Here, XAI is a powerful tool for achieving transparency and reaching ethical standards for AI. It supports interpreting, explaining, and understanding complicated AI models and their decisions. However, this application of XAI can also raise ethical concerns. In a highly competitive and rapidly developing area such as wireless communication, ethical use of XAI can become a transparency issue where it is not required. Therefore, ethical guidelines for applying XAI, processing outputs, and storing results should be transparent, fair, and accountable.

\subsubsection{Impact on B5G}
In industry 5.0, where mass personalization driven by IoT and AI \cite{ chander2022artificial } is expected to overlap with 5G and beyond future networks, XAI would be playing a vital role in ensuring security. While XAI would help localize the faults of AI models in vastly heterogeneous products and prevent hazards that could occur from external cyberattacks, the explanation tools should not be used to extract proprietary information from confidential models. Unethical extractions of data using XAI are shown to be inevitably possible in \cite{ yan2023explanation } with only query-level access to a model, which is the most common level of third-party applications' level of accessibility provided. For a security application, the risk of model extraction would be even higher, and the resulting damages would be even higher. However, if the firewalls or IDS systems fail, proprietary models deployed in the industry 5.0 are vulnerable to the same fate. Similarly, the xApps deployed in the Near-RT RIC can also be affected by a similar risk of malicious service providers stealing models from other proprietary xApps. Furthermore, with query-level access, it has been proven that XAI can create adversarial attacks that bypass malware detectors \cite{ aryal2024explainability, severi2021explanation }. 

\subsubsection{Possible solutions}
The standardization of application development using XAI for RIC or core and backhaul networks is necessary. With standardizations, organizations would implement strict access control mechanisms, advanced encryption and data masking techniques, and regular security audits to keep the unethical usage of XAI outputs in check. It is of immense significance for 6G and subsequent networks since the novel technological framework is being used in the real world for the first time. Such precautions guarantee that only authorized individuals may receive sensitive explanations derived from proprietary models and that the system is safeguarded from malevolent entities. In addition, it is essential to monitor XAI systems for the discovery of anomalies and to use differential privacy approaches to safeguard the secrecy of the underlying models. Regular security audits and penetration testing should also be conducted to identify and mitigate vulnerabilities. Ultimately, the critical aspect is the establishment of ethical principles and a framework for governance. It is essential to create a framework based on the above steps rather than a set of laws as it is more flexible, and its application can be molded to the rapidly changing landscape of AI techniques. Furthermore, it is essential to organize user education and awareness initiatives to enlighten users and developers about the ethical ramifications of XAI and the need to adhere to optimal security protocols.

\subsubsection{Summary}
In the realms of 5G and beyond networking, XAI's ethical concerns regarding accountability, transparency, and fairness are non-trivial. Although we have discussed the ethical aspects of XAI in various forms throughout the article, in this section, we have formalized the potential challenges of using XAI from an ethical perspective. Adversarial usage, proprietary data leakage, and privacy concerns are among the leading ethical concerns of security auditing in 5G and beyond networks. Also, we have proposed an ethical framework that could be used as a potential solution to establish and encourage fair usage of XAI.

\section{XAI B5G Security Projects and Standardization} \label{sec:projects}

Numerous B5G research initiatives are underway, bringing together academic and industry partners worldwide. This section summarizes several of those initiatives and their primary objectives.

\subsection{Research Projects}

1) SPATIAL \cite{SPATIAL} - EU-funded project addressing identified gaps in data and black-box AI through the design and development of resilient accountable metrics, privacy-preserving methods, verification tools, and system solutions that will serve as critical building blocks for trustworthy AI in ICT systems and cybersecurity. The project addresses the uncertainties inherent in artificial intelligence that directly impact privacy, resilience, and accountability. The SPATIAL project identifies possible XAI attacks and potential XAI technique misjudgment. As a result, it seeks to propose robust accountability metrics and integrate them into existing "black-box" AI algorithms. Another objective of the SPATIAL project is to develop detection mechanisms for detecting data biases and conducting descriptive studies on the various data quality trade-offs associated with AI-based systems.

2)CONFIDENTIAL6G - The Confidential6G project aims to improve data privacy and security in 6G networks by using sophisticated orchestration mechanisms and federated artificial intelligence/machine learning inside confidential computing frameworks. The project uses advanced blockchain technology for data verification and access control. This will be further improved by incorporating cryptographic tools where they target to develop an extensive set of tools for safeguarding privacy and implementing post-quantum cryptography. These tools specifically cater to the requirements of 6G, focusing on secure computing and networking. The project also expands on encryption and secure multi-party computing specifically designed for efficient use in collaborative AI applications and IoT edge situations in the 6G. XAI plays a vital role in spotting potential poisoning threats, comprehending their effects on local FL models, and safeguarding data privacy. This comprehensive strategy guarantees that 6G networks are well-prepared to address the crucial issues of data privacy and security.

3) 6G-GOALS \cite{6G-GOALS} - 6G-GOALS is a revolutionary initiative aimed at improving wireless system design by focusing on the significance, relevance, and value of transmitted data. The project aims to reduce data traffic by conveying only the most relevant information and designing data-efficient, robust, and resilient protocols using modern AI/ML techniques. The 6G-GOALS research breakthroughs include developing AI/ML-empowered semantic data representation, sensing, and compression algorithms, combining data-and-model-driven approaches. It also introduces semantic-oriented solutions for supporting distributed reasoning and time-sensitive communication, generalizing the low latency of 5G by tailoring communication to the actual goal. Additionally, 6G-GOALS introduces wireless technologies for sustainability in energy efficiency, EMF exposure, and spectrum management, defining the concept of semantic cognitive radio. The project aims to exploit untapped gains from AI-based joint source-channel coding and adapt to network conditions and communication objectives.

4) Hexa-X I \& II \cite{HexaX, HexaXII}- Hexa-X is an EU-funded project that aims to create the foundation for an end-to-end system architecture for 6G covering multiple fronts, including intelligent connections and radio performance. The project plans to leverage cognition, synched bio, sustainability, real-time control, and trustworthiness. Their work packages 4 and 6 focus mainly on AI-driven communication and intelligent orchestration and management of future networks. FED-XAI proposed in the project targets improving the user experience by helping the end-user trust the decisions performed by in-network AI. All the innovations led by the project are directed at mobile operators as the primary beneficiaries in the market. However, FED-XAI also targets the OEM (Original Equipment Manufacturer), enabling them to provide novel services and expand the mobile network optimization market share. In addition, Hexa-X brings forth other innovations, such as AI MANO and Zero-Energy devices, which are being collaborated with by multiple partners across Europe. On the other hand, Hexa-X-II aims to advance 6G research by developing proof-of-concepts and integrating a comprehensive 6G platform for service delivery. Additionally, Hexa-X-II seeks to influence the global 6G roadmap through impactful standardization activities. 

5) ROBUST6G \cite{robust6G} - The ROBUST-6G project focuses on the development of data-driven, AI/ML-based security solutions, addressing the evolving challenges in the forthcoming 6G services and networks. The project aims to advance security measures while safeguarding the integrity of AI/ML systems from potential breaches and upholding the privacy rights of individuals connected to the system. Furthermore, the ROBUST-6G initiative promotes green and sustainable AI/ML methodologies (including XAI), aiming to optimize energy efficiency in 6G network design. Project objectives include creating robust and sustainable approaches for AI-powered security features and making them energy-efficient, transparent, high-performing, and capable of safeguarding privacy; attaining automated, hands-free security and resource management, which enables reliable and certified services to stakeholders. More importantly, the project targets using XAI to identify and counteract threats on both network infrastructure and user devices. This involves suggesting innovative physical layer security strategies specifically designed for low latency and low energy-consuming scenarios.

6) 6G Flagship \cite{6GFLAGSHIP} - is a research project funded by the Academy of Finland that aims to commercialize 5G networks and develop a new 6G standard for future digital societies. 6G Flagship's primary objective is to develop the fundamental techniques required to enable 6G. The 6G Flagship research program recently published the world's first 6G white paper \cite{6GflagshipWhitepaper}, paving the way for the definition of the wireless era in 2030. The authors of that paper identified several intriguing security challenges and research questions, including how to improve information security, privacy, and reliability via physical layer technologies and whether this can be accomplished using quantum key distribution. Additionally, the 6G Flagship project will focus on key technology components of 6G mobile networks, including wireless connectivity, distributed intelligent computing, and privacy. Finally, with the support of industry and academia, the 6G flagship project will conduct large-scale pilots with a test network.

7) iTrust6G \cite{itrust6g} - The iTrust6G project presents a software-defined security architecture for 6G networks that follows the zero-trust principle. This architecture is built around key elements such as zero-trust AI/ML to detect and predict anomalies and threats, continuous monitoring and threat assessment, implementation of explainable intent-base E2E security orchestration using AI/ML, and improved observability in secure multi-tenancy support. The iTrust6G architecture aims to tackle challenges such as the lack of confidence in 6G platform providers in services managed by their operators in edge hardware, as well as the safe administration of various resources, which necessitates a re-evaluation of established trust approaches. The system utilizes AI and ML techniques to determine trust levels based on various metrics. In addition, iTrust6G works towards guaranteeing uninterrupted service throughout 6G network operations by using sophisticated end-to-end security orchestration techniques that are easily accessible and explainable.

8) INSPIRE-5Gplus \cite{NSprojects-Inspire5GPlus} -  the project aims to advance the security and privacy of 5G and Beyond networks. Grounded in an integrated network management system and relevant frameworks, INSPIRE-5Gplus is devoted to improving security at various dimensions, i.e., overall vision, use cases, architecture, integration to network management, assets, and models. INSPIRE-5Gplus addresses key security challenges through vertical applications ranging from autonomous and connected cars to Critical Industry 4.0. INSPIRE-5Gplus will devise and implement a fully automated end-to-end smart network and service security management framework that empowers protection, trustworthiness, and liability in managing 5G network infrastructures across multi-domains. The conceptual architecture of INSPIRE-5Gplus is split into security management domains (SDM) to support the separation of security management concerns. Each SMD is responsible for intelligent security automation of resources and services within its scope. The end-to-end (E2E) service SMD is a special SMD that manages the security of end-to-end services. The E2E service SMD coordinates between domains using orchestration. Each SMD, including the E2E service SMD, comprises a set of functional modules that operate in an intelligent closed-loop way to provide software-defined security orchestration and management that enforces and controls security policies of network resources and services in real-time.

\subsection{Standarization related to AI security}

Standardization is critical for defining the technological requirements for B5G networks and should be utilized to determine the most appropriate technologies for 6G network deployment. Thus, standards shape the global telecommunications marketplace. Numerous Standards Developing Organizations (SDOs) are tasked with standardizing 6G. Table \ref{table:specs} summarizes standardization activities in artificial intelligence security.

No standardization techniques are specifically dedicated to addressing the application of XAI in the B5G era. Few standardized publications partially reference XAI, which is essential since the lack of explainability doubts the trustworthiness and practicality of AI/ML-based security solutions. With proper standards for explanations, the intelligibility and interoperability of systems improve by allowing resilient and accountable communication. For example, the working group IEEE XAI WG P2976\textsuperscript{\texttrademark} - Standard for XAI \cite{IEEEXAIP2976WG} establish necessary and optional criteria and limits for AI methods, algorithms, applications, and systems to be explainable in a generic sense to all AI/ML applications. On the other hand, IEEE standard on an architectural framework for XAI P2894\textsuperscript{\texttrademark} \cite{IEEEXAIP2894WG} provides the definitions, taxonomy, applications, and performance evaluation guidance for using XAI.

\begin{table*}[h!]
	\renewcommand{\arraystretch}{1.3}
	\caption{Recent important standardization efforts related to Industry AI Security
	}
	\label{table:specs}
	\begin{tabular}{|p{1.5cm}|p{4cm}|p{9cm}|p{1.5cm}|}
		\hline
		\textbf{SDO} & \textbf{Standard title} &\ \textbf{Topics/Description} & \textbf{Publication date\footnotemark}\\
				\hline \hline
		      
		        ETSI \cite{ETSITR1033055} &  ETSI TR 103 305-5 V1.1.1: Critical Security Controls for Effective Cyber Defence & The document is an evolving repository for privacy enhancing implementations using the Critical Security Controls. These presently include a privacy impact assessment and use of the Controls to help meet provisions of the EU \gls{GDPR} & September 2019\\
		        \hline
		        
		        ETSI \cite{ETSIGRSAI004}&  ETSI GR SAI 004 V1.1.1: Securing Artificial Intelligence (SAI) & The standard covers the problem of securing AI-based systems and solutions, with a focus on machine learning, and the challenges relating to confidentiality, integrity and availability at each stage of the machine learning lifecycle.  & December 2020\\
		        \hline
		        
		        ETSI \cite{ETSITR103674}&  ETSI TR 103 674 V1.1.1: Artificial Intelligence and the oneM2M architecture & The ETSI TR 103 674 is addressing the issues related to the introduction of AI into \gls{IoT} systems and, as first priority, into the oneM2M architecture.  & February 2021\\
		        \hline
		        
		        ETSI \cite{ETSITR103675}&  ETSI TR 103 675 V1.1.1: SmartM2M; AI for IoT: A Proof of Concept & This standard covers the implementation and security challenges of oneM2M platforms using the case of \gls{IoT}. It shows how the ML methods can be implemented directly over the data. & December 2020\\
		        \hline
		        
		        ETSI \cite{ETSIGRZSM010}&  ETSI GR ZSM 010 V1.1.1: Zero-touch network and Service Management (ZSM); General Security Aspects & The document studies the security aspects of the ZSM use cases, framework and solutions, identifies potential security threats and mitigation considerations to be covered in ZSM standardization activities. It aims to outline a list of security controls (aka security countermeasures) in order to raise awareness of security aspects that could be considered in ZSM specifications. The cited document explores the relationship between security controls and technology-specific solutions. & July 2021\\
		        \hline
		       
		        ISO \cite{ISOTR24028}&  ISO/IEC TR 24028:2020: Information technology — Artificial intelligence — Overview of trustworthiness in artificial intelligence & This document surveys approaches to establish trust in AI systems through transparency, explainability, controllability. It also surveys engineering pitfalls and typical associated threats and risks to AI systems, along with possible mitigation techniques and methods. & May 2020\\
		        \hline
		        
		        NIST \cite{NISTIR8228}&  NISTIR 8228: Considerations for Managing \gls{IoT} Cybersecurity and Privacy Risks & The purpose of NISTIR 8228 is to help federal agencies and other organizations better understand and manage the cybersecurity and privacy risks associated with their individual \gls{IoT} devices throughout the devices’ lifecycles.  & June 2019 \\
		        \hline
		        
		        NIST \cite{NISTSP80053} &  NIST SP 800-53 revision 5: Security and Privacy Controls for Federal Information Systems and Organizations & NIST SP 800-53 is designed to help organizations identify the security and privacy controls needed to manage risk and to satisfy the security and privacy requirements. It accomplishes this objective by providing a comprehensive and flexible catalog of security and privacy controls to meet current and future protection needs based on changing threats, vulnerabilities, requirements, and technologies. & September 2020\\
		        \hline
		        
		        IEEE \cite{IEEP7001}&  IEEE P7001: Transparency of autonomous systems & This standard describes measurable, testable levels of transparency, so that autonomous systems can be objectively assessed and levels of compliance determined. & June 2020\\
		        \hline

		        IEEE \cite{IEEP7006}& IEEE P7006: Personal data AI agent (working group) & This working group works on a standard that describes the technical elements required to create and get access to a personalized (AI) that will comprise inputs, learning, ethics, rules and values controlled by individuals. & June 2021\\
		        \hline
		       
	\end{tabular}
	
\end{table*}
\vspace{1em}

\section{Lessons Learned and Future Research Directions} \label{sec:lesson}

This section discusses the lessons learned. Based on these lessons we synthesize the future research directions that industrial or academic researchers can follow.

\begin{figure*}[htb]
    \centering
    \includegraphics[width=0.98\textwidth]{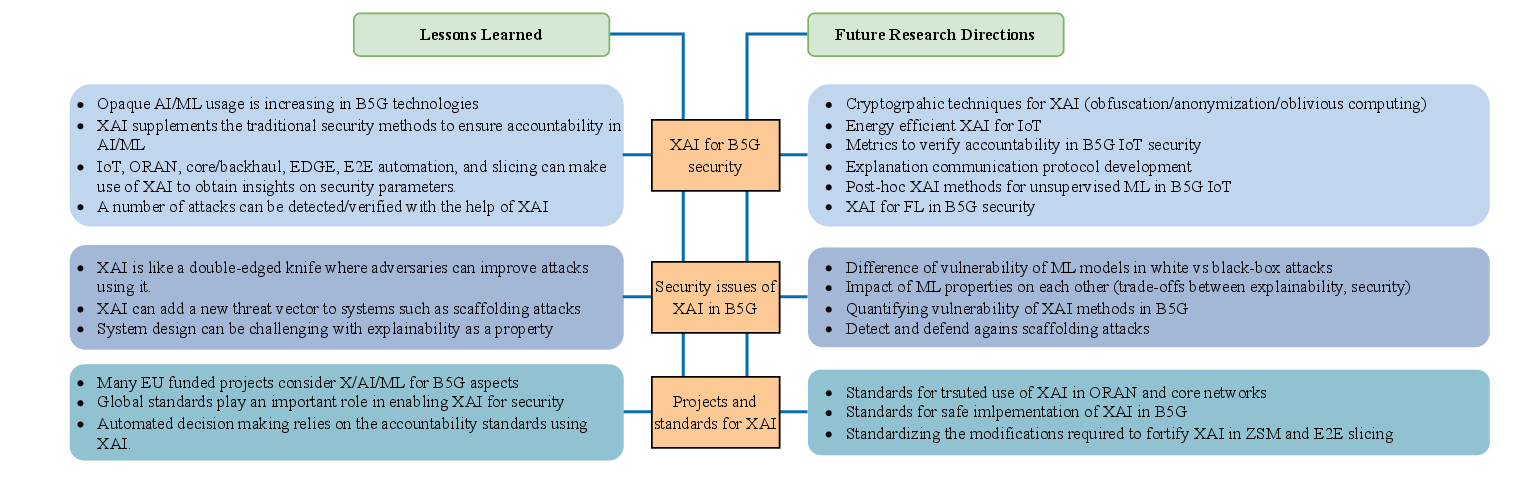}
    \caption{Summary of future research directions and lessons learned}
    \label{fig:futureresearch}
\end{figure*}

\subsection{Role of XAI for B5G Security} 
\subsubsection{Lessons Learned}

B5G and future telecom infrastructure will include ML. The opaque decision-making process of AI/ML methods raises concerns about security responsibility. This limitation prevents many essential B5G service providers from utilizing AI/ML in real-world applications. 
IoT enables B5G. Due to close physical access to users, deployed AI/ML models can expect increased adversarial manipulations. Jamming, side-channel attacks, botnets, MITM, and DDoS attacks are a few. AI/ML methods are widely used to detect attacks, but their trustworthiness is questionable for critical applications. LIME/SHAP and other explanation generation methods may provide clarity and accountability when realizing the impact of security parameters on IoT devices. It could massively impact stakeholder decision-making. V/C/ORAN technologies shape B5G radio infrastructure depending on AI/ML outputs (e.g., SMO RICs). Our survey sheds light on the fact that trusted and accountable XAI could improve the security of ORAN's AI/ML workflow by bridging operators' understanding of ML models and adversarial actions.
On the other hand, edge AI is growing in popularity across industries due to its affordability and data privacy. Edge AI will power services, including health monitoring and military applications. Predictable ML model behaviors expedite the creation of security and transparency of edge protocols. To conclude, the ultimate objective of B5G with end-to-end slicing and closed-loop automation is enabled with trustworthy AI. A transparent security architecture for network management is pertinent in the face of the growth of adversarial ML methods. In this regard, architects should provide an interpretable channel for supervised and unsupervised learning methods. The number of XAI methods available for supervised learning methods is vastly higher than unsupervised ones. Also, the stability, comparability, efficiency, and accuracy of these XAI methods could be improved further. Since federated learning is one of the leading technologies in privacy-preserving B5G, the application of XAI has become a growing need in this sector.

\subsubsection{Open Research Challenges}

\begin{itemize}
    \item How to modify cryptography techniques (obfuscation/anonymization/oblivious computing) to complement XAI or vice versa?
    \item How to develop energy-efficient XAI methods for low-power IoT devices?
    \item How to measure the accuracy/accountability of explanations for ORAN (RIC functions) and Core NFs?
    \item How to develop safe protocols to deliver explanations to relevant stakeholders without losing their integrity?
    \item How to apply reliable post-hoc XAI methods to unsupervised/reinforcement machine learning techniques in slice resource optimization?
    \item How to safely apply post-hoc XAI methods in federated learning techniques without compromising privacy? 
\end{itemize}

Developing energy-efficient XAI techniques is a primary requirement for low-powered IoT devices. The metrics must also adhere to the computational efficiency criteria. For encryption and decryption tasks, extra computational power will be a requirement to run alongside security measures like post-hoc explainers. It will further limit the computation power to improve service accountability and trustworthiness. Popular XAI methods such as SHAP are computationally expensive when run iteratively. An extensive research gap remains for energy-efficient XAI methods. Quantitative analysis of explanations is essential for E2E network automation. Explanations generated in different slices require analysis without compromising inter-slice isolation. Also, RL methods used in resource allocation are not backed with explanations for resilience and accountability \cite{cheng2023reinforcement}. 

\subsubsection{Preliminary solutions}

Ongoing work is currently to use GPU-based acceleration for post-hoc explainers \cite{nvidia2022xaisynthdata}. However, it is currently not supported fully for popular XAI methods such as SHAP (only available for TreeSHAP). With dedicated accountability provided through GPUs, supplementary security methods such as encryption techniques can use the remaining computation power. There are multiple attempts to introduce metrics for interpretability, but it remains a heavily domain-subjective concept and an open research question \cite{hoffman2018metrics}. Some studies propose metrics \cite{li2020trustworthy} to quantify the quality of ML model interpretations mathematically. However, generic metrics are insufficient when applied for NFs in the core and ORAN(RIC) functions. This is an open research question that requires immediate attention at the moment.
The current post-hoc XAI methods have limited compatibility in unsupervised learning techniques \cite{morichetta2019explain, wickramasinghe2021explainable}. Therefore, such techniques must be further analyzed and improved for industrial standards. Some work, such as \cite{renda2022federated}, lays the foundation for XAI in FL for 5G and beyond systems. However, they are limited to FL models that are not based on the optimization of differentiable global objective functions. 

\subsubsection{Possible Future Directions}

Interpreters must be carefully adjusted to filter out any sensitive information generated to avoid privacy violations and intellectual property laws before conveying them to the stakeholders. Here, we set forth the importance of testing XAI alongside privacy-preserving methods such as differential privacy and data anonymization techniques. 
Oblivious computing is an up-and-coming security method that adds high protection for user data. Under this umbrella, it would be interesting research to see the possibilities of applying XAI methods alongside homomorphic encryption and multi-party computation \cite{mohassel2017secureml, kitai2019mobius, gilad2016cryptonets}.
The development of new protocols will require compliance with legal frameworks and standards. It should be accompanied by extensive research on developing security metrics to quantify and detect problems in explanations. These metrics can then be replicated for fully automated XAI network management (ZSM). XAI in quantum networks is highly preliminary, and further research is required to apply it to parameterized quantum circuits. Work such as \cite{steinmuller2022explainable} must be extended for XAI methods other than SHAP and integrated gradients for an overall understanding of the domain. Enabling this functionality will require novel protocols to support unsupervised/reinforcement learning techniques. 
XAI methods for FL models are currently actively pursued in research. It would be interesting to see how XAI can improve the accountability of FL in distributed ORAN/core resource allocation methods without compromising their privacy.

\subsection{Research projects and standards}
\subsubsection{Lessons Learned}

According to our research, several EU-funded research projects have already started to address the challenges on the path toward 6G, and many major ICT companies are issuing announcements about internal programs focusing on 6G security. Outside of the EU, e.g., in the USA, the Next G Alliance started to work on the 6G security and privacy through private sector-led efforts. Most of the projects listed in Section \ref{sec:projects} aim to guarantee the following generation network's trustworthiness and security. However, it is exciting to see approaches beyond classical, for example, XAI-based techniques, to secure future networks that play a significant role in most of the research projects reviewed in this paper. Undoubtedly, global standards and new regulations will play a key role in developing and deploying 6G networks. However, effective and timely standardization is key to the fast and seamless adoption of new technologies, including 6G. Several Standards Developing Organizations (SDO) are expected to work in the near future or already work on 6G security and privacy, e.g., ETSI, IETF, IEEE, 3GPP, NIST, and ISO, in a much tighter way than they did for 5G, as 6G aims to merge different technologies already standardized by SDOs. The AI/ML mechanisms will have to become the main elements in 6G to achieve superior security, e.g., automating decision-making processes and accomplishing a zero-touch approach. 

\subsubsection{Open Research Challenges}

\begin{itemize}
    \item How are the new standards needed to be prepared for the safe application of XAI in ORAN and the core network?
    \item How to implement standards that are required to ensure the security of XAI in the 5G and beyond era?
    \item How to modify the standards of ZSM and E2E slicing security to add the layer of accountability through XAI? 
\end{itemize}

The current need for standards for XAI methods is astonishingly low. Research on the security of XAI in enablers such as core and ORAN has been opening a significant research gap for standards and protocols lately. ETSI architecture is currently not accommodating XAI in the ZSM and E2E slicing but has emphasized the importance of XAI. This gives a significant research challenge for further explorations.

\subsubsection{Preliminary Solutions}
The analysis of recently released standards (2019-2021) in B5G security shows that most SDOs acknowledge the importance of AI/ML-based security solutions for B5G networks. However, only a few standardization documents mention the role of XAI, which is very significant, as the current lack of explainability leads to doubts about the credibility and feasibility of AI/ML-based implementations built to combat security threats. There are, however, working groups, such as IEEE XAI WG - Standard for XAI \cite{IEEEXAIP2976WG} that aim to standardize mandatory and optional requirements and constraints that need to be satisfied for AI methods, algorithms, applications, or systems to be recognized as explainable.

\subsubsection{Possible Future Directions}

The composition of more meticulous standards on the elements of XAI security and its provision of transparent AI/ML techniques for B5G security is a requirement. The European Partnership on Smart Networks and Services (SNS) established Europe's strategic research and innovation roadmap. The initiative is based on an EU contribution of €900 million over the next seven years. The objective is to enable European players to develop R\&I capabilities for 6G systems and lead markets for 5G and 6G infrastructure, which will serve as the foundation for digital and green transformation. The SNS work program will be the basis for calls for proposals aimed to launch in early 2022. Concerning standards, we believe that projects under calls such as ICT-52-2020 expect to provide valuable inputs to standardization bodies fostering the development of advanced 6G solutions. From the perspective of 3GPP, there are features and capabilities from existing 5G solutions that require full specification and are expected to be released at the end of 2023. The migration from legacy and existing proprietary radio protocols toward 3GPP protocols will take 5-10 years. AI/ML-assisted security still needs further development to respond to new security threats introduced by the dynamicity of 6G services and networks.

\section{Conclusion}\label{sec:conclusion}

This survey examines and evaluates the potential of using XAI to improve accountability and resilience beyond the 5G era of AI-based security in communication. The study begins by laying the background of current XAI technical concepts and their potential in the B5G era. This paper discussed an exhaustive assessment of the most cutting-edge AI, XAI, B5G technologies, and security aspects, including threat models and taxonomy. Technical aspects regarding the role of XAI in B5G security issues were thoroughly examined in three main layers of the B5G era. Here, we discuss enablers such as IoT, RAN, Edge, core, backhaul, E2E slicing, and network automation. We also discuss how XAI can be associated with security mechanisms such as encryption, anonymization, obfuscation, and federated learning. It is followed by a detailed discussion on trending AI-based use cases of B5G and XAI's potential in ensuring those networks' trustworthiness. Apart from the favorable prospects of XAI, we also bring to light new security issues and challenges introduced to future network infrastructure along with AI explanations. 
Later in this paper, we focus on the active research initiatives to build and standardize B5G-specific technologies involving researchers and industry practitioners. Finally, this paper highlights lessons learned and future research directions for readers to pursue. In conclusion, this survey acts as a stepping stone for researchers, industry partners, or other stakeholders to absorb a holistic understanding of the potential of XAI to improve accountability and resilience in the security application of the B5G era.

\section*{Acknowledgement}
This work is partly supported by the European Union in the SPATIAL Project (Grant No: 101021808), Confidential-6G (Grant No: 101096435), Robust-6G (Grant No: 101139068) and Science Foundation Ireland under the CONNECT phase 2 (Grant no. 13/RC/2077\_P2) projects. We would also like to thank Zujany Salazar for her contribution to the early version of this paper when she was with Montimage.

\bibliographystyle{IEEEtran}
\bibliography{references.bib}

\begin{IEEEbiography}[{\includegraphics[width=1in,height=1.25in,clip,keepaspectratio]{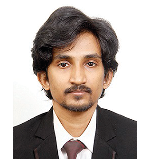}}]{Thulitha Senevirathna} received the degree for bachelor of the science in electrical and information engineering from University of Ruhuna, Sri Lanka. He is currently pursuing the PhD degree affiliated to the school of computer science in University College Dublin. His research interests include machine learning, explainable AI and AI security in B5G applications.
\end{IEEEbiography}


\begin{IEEEbiography}[{\includegraphics[width=1in,height=1.25in,clip,keepaspectratio]{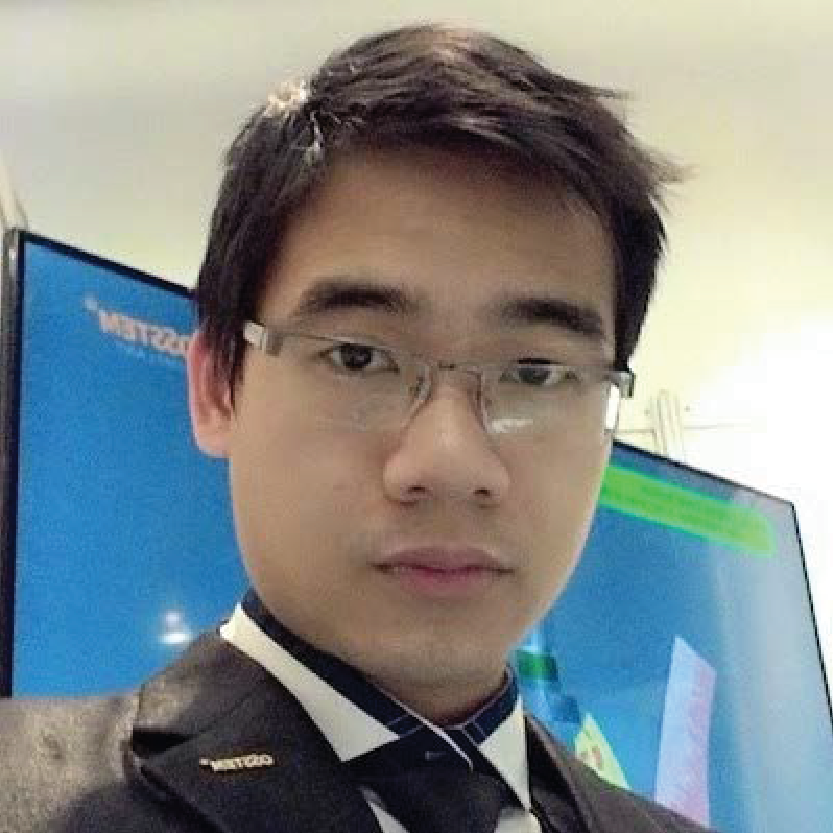}}]{Vinh Hoa La} is currently a R\&D Engineer and a project manager for the EU H2020 SPATIAL project at Montimage, an innovative company located in Paris. He received his engineering degree in Information and Communication Systems at Hanoi University of Science and Technology (Vietnam) in 2012 and the Master degree at UPMC-Paris 6 in 2013. In 2016, he was honored the title Doctorate of Telecom SudParis /Univeristy Paris Saclay. His research interests include Security Monitoring, 5G/IoT/Sensors Network Security and Root-cause Analysis.
\end{IEEEbiography}

\begin{IEEEbiography}[{\includegraphics[width=1in,height=1.25in,clip,keepaspectratio]{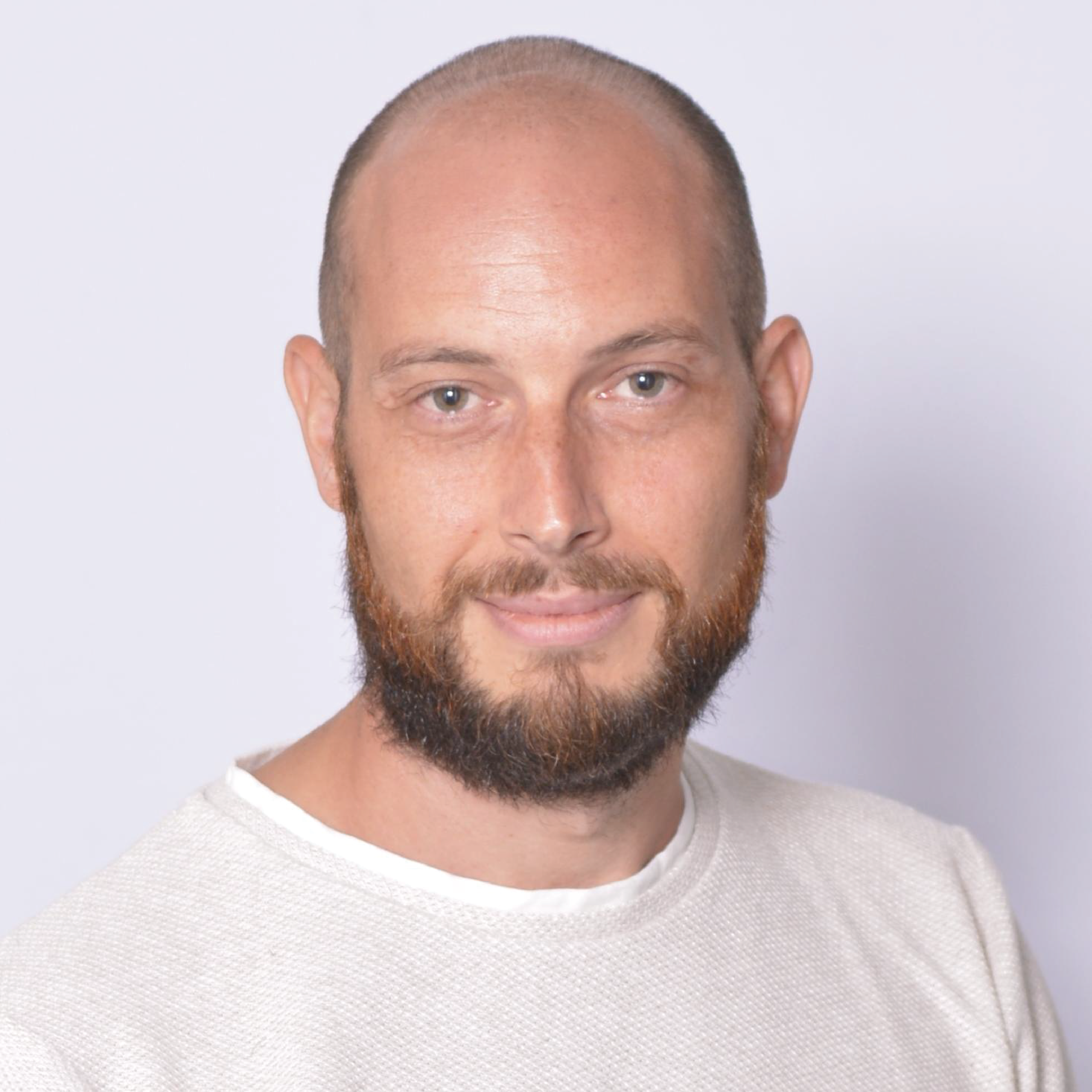}}]{Samuel Marchal} Samuel Marchal is the Research Team Leader for the Network Security research group at VTT Technical Research Centre of Finland. He received the engineer's and M.Sc. degree from TELECOM Nancy, France, and the Ph.D. degree from the University of Luxembourg. Samuel has over 12 years of academic research experience in system security, network security and AI security. He has led and been part of several collaborative research projects with organizations such as Intel, Huawei, McAfee or Zalando. More recently, he led the research on AI security at F-Secure and then at WithSecure, developing approaches for the security assessment of AI systems. His latest research interests lie at the intersection of AI/ML and cybersecurity, namely using AI for cybersecurity, the security of AI systems and the use of AI in cyberattacks.
\end{IEEEbiography}

\begin{IEEEbiography}[{\includegraphics[width=1in,height=1.25in,clip,keepaspectratio]{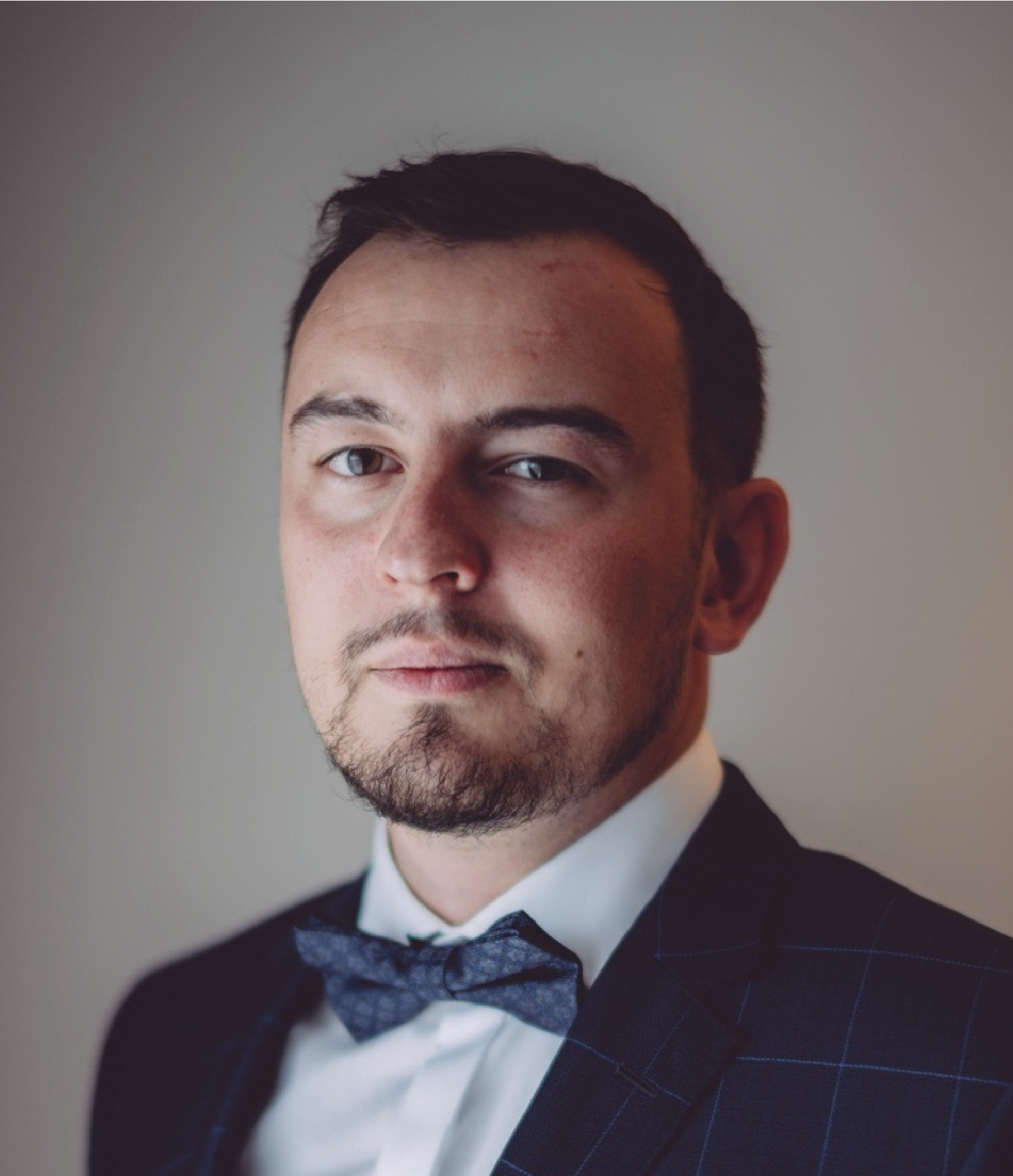}}]{Bartlomiej Siniarski}(Member, IEEE) is currently a post doctoral researcher and a project manager for the EU H2020 SPATIAL project at University Colelge Dublin. He completed his undergraduate studies in Computer Science at University College Dublin (Ireland) and University of New South Wales (Australia). He was awarded with a doctoral degree in 2018. He has a particular interest and experience in the design of the \gls{IoT} networks and in particular collecting, storing and analysing data gathered from intelligent sensors. Furthermore, he was actively involved in MSCA-ITN-ETN, ICT-52-2020 and H2020-SU-DS-2020 projects which are focused on solving problems in the area of network security, performance and management in 5G and B5G networks.
\end{IEEEbiography}

\begin{IEEEbiography}[{\includegraphics[width=1in,height=1.25in,clip,keepaspectratio]{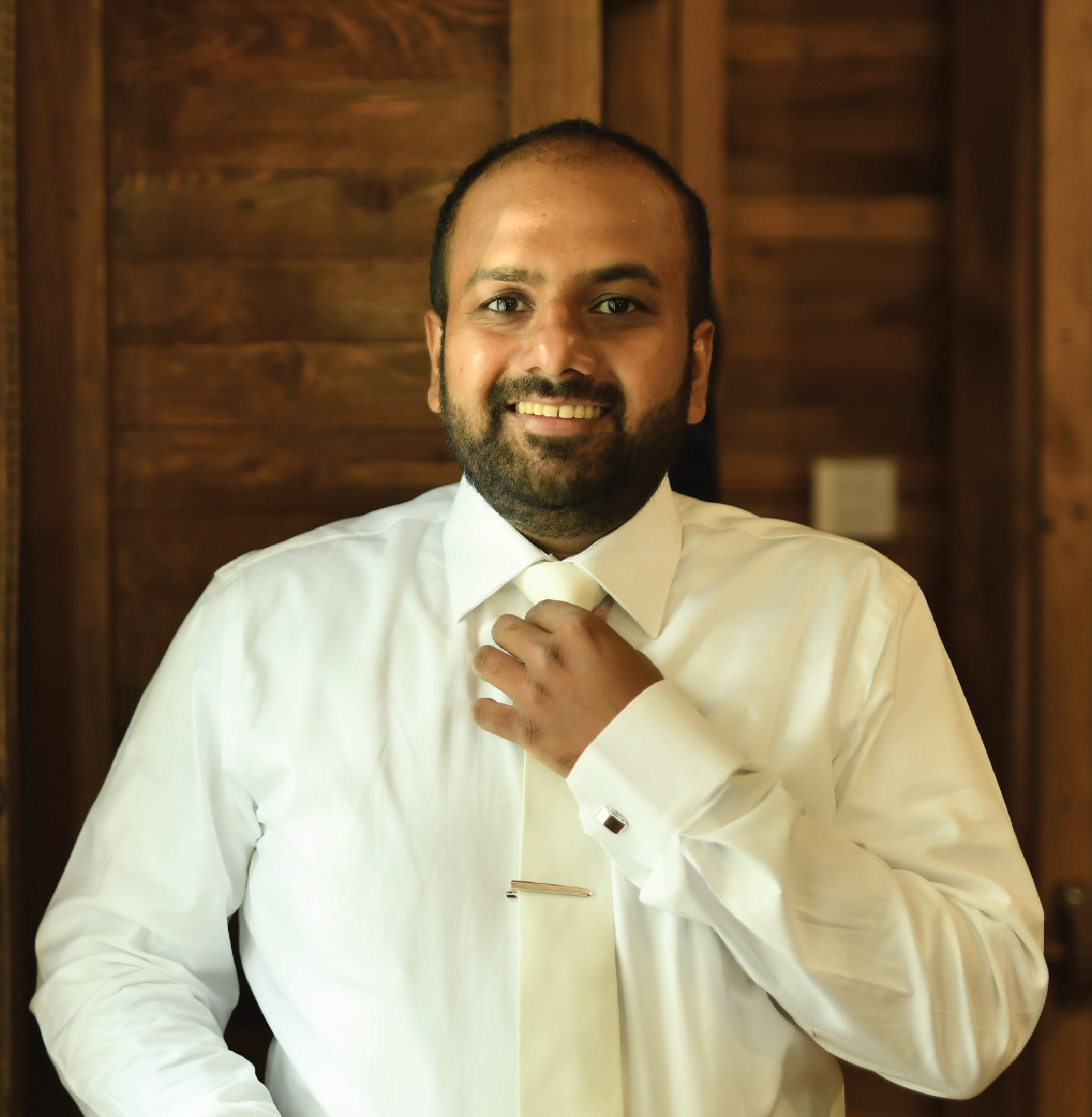}}]{Madhusanka Liyanage} (Senior Member, IEEE)  is an Associate Professor/Ad Astra Fellow and Director of Graduate Research at the School of Computer Science, University College Dublin, Ireland. He is also the director of Network Softwarization and Security Labs (NetsLab) at the UCD School of Computer Science. Moreover,  He is an Honorary Adjunct Professor at the University of Ruhuna, Sri Lanka and the University of Sri Jayawardhanepura, Sri Lanka. He received his Doctor of Technology degree in communication engineering from the University of Oulu, Oulu, Finland, in 2016. He was also a recipient of the prestigious Marie Skłodowska-Curie Actions Individual Fellowship and the Government of Ireland Postdoctoral Fellowship during 2018-2020.  In 2020, he received the "2020 IEEE ComSoc Outstanding Young Researcher" award by IEEE ComSoc EMEA. In 2021, 2022 and 2023, he was ranked among the World's Top 2\% Scientists (2020, 2021 and 2022) in the List prepared by Elsevier BV, Stanford University, USA. Also, he was awarded an Irish Research Council (IRC) Research Ally Prize as part of the IRC Researcher of the Year 2021 and 2023 awards for the positive impact he has made as a supervisor.  In 2022, he received "2022 The Tom Brazil Excellence in Research Award" by SFI CONNECT Center.  Dr. Liyanage's research interests are 5G/6G, Blockchain, Network security, Artificial Intelligence (AI), Explainable AI, Federated Learning (FL), Network Slicing, Internet of Things (IoT) and Multi-access Edge Computing (MEC).  More info: \url{www.madhusanka.com} 
\end{IEEEbiography}

\begin{IEEEbiography}[{\includegraphics[width=1in,height=1.25in,clip,keepaspectratio]{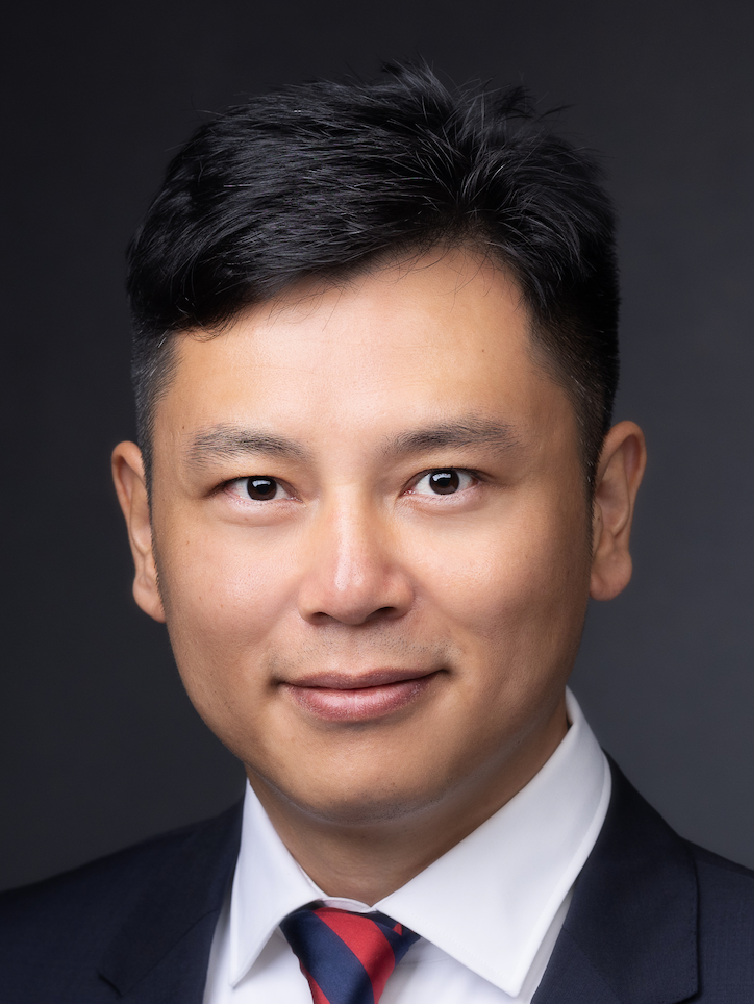}}]{Shen Wang}
(Senior Member, IEEE) received the M.Eng. degree from Wuhan University, China, and the Ph.D. degree from Dublin City University, Ireland. He is currently an Assistant Professor with the School of Computer Science, University College Dublin, Ireland. He has been involved with several EU projects as a co-PI, the WP, and a task leader of big trajectory data streaming for air traffic control and trustworthy AI for intelligent cybersecurity systems. Some key industry partners of his applied research are IBM Research Brazil, Boeing Research and Technology Europe, and Huawei Ireland Research Centre. His research interests include connected autonomous vehicles, explainable artificial intelligence, and security and privacy for mobile networks.
\end{IEEEbiography}

\vfill

\end{document}